\documentclass[twocolumn,secthm]{autart}    

\usepackage{graphicx}          
\usepackage[hyphens]{url}

\usepackage[english]{babel}

\usepackage[latin1]{inputenc}
\usepackage[T1]{fontenc}

\usepackage[active]{srcltx}

\usepackage{siunitx}
	\DeclareSIUnit{\var}{var}
\usepackage{lscape}		
\usepackage{framed}		
\usepackage{longtable}
\usepackage{booktabs}	
\usepackage{fancyhdr}
\usepackage{setspace}
\usepackage{enumerate} 
\usepackage{multienum}
\usepackage[font=small,textfont=it]{caption}	
\usepackage{xcolor}
\usepackage{comment}
\usepackage[ngerman, num]{isodate}	
\usepackage{afterpage}
\usepackage{eurosym}

\usepackage{bibentry}

\usepackage{float}
\usepackage{subfig}
\usepackage{graphicx}
\usepackage{psfrag}
\usepackage[]{silence} %
\usepackage[process=auto,crop=pdfcrop,cleanup={.tex,.dvi,.ps,.pdf,.snm,.out,.bbl,.nav,.aux,.auxlock,.toc}]{pstool}

\usepackage{stmaryrd}	

\usepackage{makeidx}
\usepackage{lastpage}

\usepackage{algorithm}
\usepackage{algpseudocode}
\usepackage{forloop}

\usepackage{amsmath}

\usepackage{amssymb}
\usepackage{mathtools}
\usepackage{bm}
\usepackage{arydshln} 
\newcommand{\mat}[1]{\ensuremath{\bm{#1}}} 
\renewcommand{\vec}[1]{\ensuremath{\bm{#1}}} 
\DeclareMathOperator{\rank}{rank} 
\newcommand{\T}{\ensuremath{{\top}}} 
\newcommand{\define}{\ensuremath{\coloneqq}} 
\newcommand{\definer}{\ensuremath{=:}} 
\newcommand{\isomorph}{\ensuremath{\cong}} 
\DeclareMathOperator{\blkdiag}{blkdiag} 
\DeclareMathOperator{\Spec}{Spec} 
\DeclareMathOperator{\img}{im} 
\newcommand{\N}{\ensuremath{\mathbb{N}}} 
\newcommand{\R}{\ensuremath{\mathbb{R}}} 
\newcommand{\posdef}{\ensuremath{\succ 0 }} 
\newcommand{\possemidef}{\ensuremath{\succeq 0}} 
\newcommand{\I}{\ensuremath{\mat{I}}} 
\newcommand{\blockI}{\ensuremath{\I}} 
\newcommand{\0}{\ensuremath{\mat{0}}} 
\newcommand{\blockZero}{\ensuremath{\0}} 
\newcommand{\card}[1]{\abs{#1}} 

\newcommand{\vecspace}[1]{\ensuremath{\mathcal{#1}}} 
\newcommand{\set}[1]{\ensuremath{\mathbb{#1}}} 

\newcommand{\dirac}{\ensuremath{{\vecspace{D}}}} 
\newcommand{\X}{\ensuremath{{\vecspace{X}}}} 

\newcommand{\n}{\ensuremath{{n}}} 

\newcommand{\F}{\mat{F}}
\newcommand{\E}{\mat{E}}
\newcommand{\matZ}{\ensuremath{\matioD}} 
\newcommand{\matioD}{\ensuremath{\mat{Z}}} 

\newcommand{\matJ}{\ensuremath{{\mat{J}}}} 
\newcommand{\matR}{\ensuremath{{\mat{R}}}} 
\newcommand{\matG}{\ensuremath{{\mat{G}}}} 
\newcommand{\matP}{\ensuremath{{\mat{P}}}} 
\newcommand{\matM}{\ensuremath{{\mat{M}}}} 
\newcommand{\matS}{\ensuremath{{\mat{S}}}} 

\newcommand{\ofx}{\ensuremath{\left( \vec{x} \right)}} 
\newcommand{\Tofx}{\ensuremath{^\T \!\left( \vec{x} \right)}} 

\newcommand{\x}{\vec{x}} 
\newcommand{\vecu}{\vec{u}} 
\newcommand{\vecy}{\vec{y}} 
\newcommand{\f}{\vec{f}} 
\renewcommand{\e}{\vec{e}} 

\newcommand{\dimBond}{\ensuremath{{N}}} 
\newcommand{\numV}{\ensuremath{{n}}} 
\newcommand{\numB}{\ensuremath{{m}}} 
\newcommand{\numVC}{\ensuremath{{\numV_\bgC}}} 

\newcommand{\bgC}{\ensuremath{{\mathrm{C}}}} 
\newcommand{\bgR}{\ensuremath{{\mathrm{R}}}} 
\newcommand{\bgSe}{\ensuremath{{\mathrm{Se}}}} 
\newcommand{\bgSf}{\ensuremath{{\mathrm{Sf}}}} 
\newcommand{\bgZero}{\ensuremath{{\mathrm{0}}}} 
\newcommand{\bgOne}{\ensuremath{{\mathrm{1}}}} 
\newcommand{\bgTF}{\ensuremath{{\mathrm{TF}}}} 
\newcommand{\bgGY}{\ensuremath{{\mathrm{GY}}}} 
\newcommand{\bgI}{\ensuremath{{\mathrm{I}}}} 
\newcommand{\bgE}{\ensuremath{{\mathrm{E}}}} 
\newcommand{\bgIC}{\ensuremath{{\mathrm{IC}}}} 
\newcommand{\bgP}{\ensuremath{{\mathrm{P}}}} 

\newcommand{\V}{\ensuremath{{\set{V}}}} 
\newcommand{\B}{\ensuremath{{\set{B}}}} 
\newcommand{\Bin}{\ensuremath{\overleftarrow{\B}}} 
\newcommand{\Bout}{\ensuremath{\overrightarrow{\B}}} 
\newcommand{\BG}{\ensuremath{{\mathcal{BG}}}} 

\newcounter{ct}
\newcommand{\markdent}[1]{\forloop{ct}{0}{\value{ct} < #1}{\hspace{\algorithmicindent}}}
\newcommand{\markcomment}[1]{\Statex\markdent{#1}}
\algrenewcommand{\algorithmiccomment}[1]{\State // #1} 
\algnewcommand\True{\textbf{true}}
\algnewcommand\False{\textbf{false}}
\algnewcommand\Print{\State \textbf{print} }
\algnewcommand\Terminate{\State \textbf{terminate} }

\newcommand{\tighteq}{\ensuremath{\!= \!}} 
\newcommand{\tightplus}{\ensuremath{\!+ \!}} 
\newcommand{\phs}{PHS}
\newcommand{\phss}{PHSs}

\graphicspath{{./pics/}}

\usepackage[final]{fixme}
\fxsetup{multiuser,layout={inline},innerlayout={layout={nomarginclue,inline}}}
\fxusetheme{color}
\FXRegisterAuthor{mp}{amp}{{\setlength{\fboxsep}{2pt}\colorbox{blue}{\color{white}MP}\color{blue}}}
\FXRegisterAuthor{sp}{asp}{{\setlength{\fboxsep}{2pt}\colorbox{yellow!60!black}{\color{black}SP}\color{yellow!60!black}}}
\FXRegisterAuthor{sc}{asc}{{\setlength{\fboxsep}{2pt}\colorbox{cyan}{\color{black}SC}\color{cyan}}}
\FXRegisterAuthor{sk}{ask}{{\setlength{\fboxsep}{2pt}\colorbox{red}{\color{white}SK}\color{red}}}

\newcommand{\sca}[1]{\scnote{#1}}

\begin{document}

\begin{frontmatter}
\runtitle{Automated Generation of Explicit Port-Hamiltonian Models from Multi-Bond Graphs} 

\title{Automated Generation of Explicit Port-Hamiltonian Models from Multi-Bond Graphs\thanksref{fnauthenticity}} 
\thanks[fnauthenticity]{This paper was not presented at any conference.}
\thanks[fncoraut]{Corresponding author is M. Pfeifer. Tel. +49 (0)721 608 43236. Fax +49 (0)721 608 42707.}

\author[IRS]{Martin Pfeifer\thanksref{fncoraut}}\ead{martin.pfeifer@kit.edu},    		
\author[IAG]{Sven Caspart}\ead{sven.caspart@kit.edu},               
\author[IRS]{Silja Pfeiffer}\ead{usdve@student.kit.edu},  
\author[IRS]{Charles Muller}\ead{ubese@student.kit.edu},  
\author[IRS]{Stefan Krebs}\ead{stefan.krebs@kit.edu},  
\author[IRS]{S\"oren Hohmann}\ead{soeren.hohmann@kit.edu},  

\address[IRS]{Institute of Control Systems (IRS), Karlsruhe Institute of Technology (KIT), Kaiserstr. 12, 76131 Karlsruhe, Germany}  
\address[IAG]{Institute of Algebra and Geometry (IAG), Karlsruhe Institute of Technology (KIT), Kaiserstr. 12, 76131 Karlsruhe, Germany}             


\begin{abstract}                          
Port-Hamiltonian system theory is a well-known framework for the control of complex physical systems. 
The majority of port-Hamiltonian control design methods base on an \emph{explicit} input-state-output port-Hamiltonian model for the system under consideration. 
However in the literature, little effort has been made towards a systematic, automatable derivation of such explicit models. 
In this paper, we present a constructive, formally rigorous method for an explicit port-Hamiltonian formulation of multi-bond graphs. 
Two conditions, one necessary and one sufficient, for the existence of an explicit port-Hamiltonian formulation of a multi-bond graph are given. 
We summarise our approach in a fully automated algorithm of which we provide an exemplary implementation along with this publication. 
The theoretical and practical results are illustrated through an academic example. 
\end{abstract}

\end{frontmatter}

\section{Introduction} \label{sec:intro}
\textbf{Motivation:}
%
The theory of port-Hamiltonian systems~(\phss) is a well-known framework for controller and observer design in complex \emph{physical} systems. 
\phs~have first been introduced for real-valued, continuous-time nonlinear systems with lumped parameters. Amongst others, pioneering works are \cite{maschke1992a,maschke1992b,vanderSchaft1995}.
Meanwhile, the port-Hamiltonian framework has been extended to complex-valued systems~\cite{mehl2016}, discrete-time systems~\cite{kotyczka2018}, and distributed-parameter systems~\cite{vanderSchaft2002,legorrec2005,jacob2012,ramirez2017}.
Port-Hamiltonian methods feature a high degree of modularity and physical insight~\cite{kugi2001,duindam2009} and have significant potential for automated control design~\cite{kotyczka2013}.
Port-Hamiltonian controller and observer design methods are model-based; 
the majority of methods base on an explicit input-state-output port-Hamiltonian model for the system under consideration, see e.g. \cite{ortega2008,doerfler2009,venkatraman2010,vincent2016,van_der_schaft2016}. 
This raises the question how such models can be derived systematically, especially for systems of high complexity.
However, as will be seen in the next paragraph there exist only few studies in the literature which address this question.  \\
In this paper we present an automatable method for the derivation of explicit input-state-output port-Hamiltonian models from multi-bond graphs. 
As multi-bond graphs are graphical system descriptions, our method allows for a comfortable and time-efficient modelling of complex physical systems.
We focus on real-valued, continuous-time, finite-dimensional \phss. 

\textbf{Related literature:}
In literature, different graphical system descriptions have been used for deriving port-Hamiltonian models of complex physical systems.
In \cite{vanderSchaftMaschke2013}, various complex systems are described as open directed graphs. Based on the graph description, explicit port-Hamiltonian models can be obtained. 
The authors of \cite{falaize2016} propose a method for the automated generation of differential-algebraic port-Hamiltonian models from schematics of analog circuits. The method is implemented in a corresponding Python tool~\cite{pyphs2019} which allows for an automated equation generation. \\
Besides directed graphs and schematics, bond graphs are a natural starting point for the derivation of explicit \phss~as both -- bond graphs and \phss~-- share an energy- and port-based modelling philosophy. 
Ref.~\cite{rosenberg1971} was the first to systematically derive a \emph{state-space} formulation of single-bond graphs.
The method is based on a mathematical representation of the bond-graph referred to as \emph{field representation} (cf. \cite[p.~220]{wellstead1979}).
%
The formulation of a single-bond graph as a \phs~was first investigated in~\cite{golo2003}. The authors show that each well-posed bond graph permits an \emph{implicit} port-Hamiltonian formulation. 
Such an implicit \phs~aims at a use in numerical simulations. For the design of control methods, however, an explicit \phs~is required. The transfer from an implicit to an explicit port-Hamiltonian representation is non-trivial. 
In particular, as we will show later, the existence of an explicit port-Hamiltonian formulation of a bond graph is not guaranteed, even if the bond graph is well-posed. \\
The author of~\cite{lopes2016} addresses the formulation of a single-bond graph as differential-algebraic \phs.
It has been shown that such a differential-algebraic \phs~can possibly be transferred into an explicit input-state-output \phs~\cite{lopes2016}.
Concerning this transfer, there exists a \emph{sufficient} condition which, however, is restrictive as it demands some block matrices of the underlying Dirac structure to be zero. 
A \emph{necessary} condition for the existence of an explicit port-Hamiltonian formulation of a bond graph is missing in the literature. 
Ref.~\cite{donaire2009} provides a method transferring a class of causal single-bond graphs to an explicit input-state-output \phs.
The approach is restricted to non-feedthrough systems.
As in \cite{rosenberg1971}, the starting point of \cite{donaire2009} is a bond graph field representation. In the field representation, the authors assume some of the block matrices to be constant or zero.
The author of \cite{dai2016} proposes a concept for formulating single-bond graphs as simulation models with port-Hamiltonian dynamics.
However, the models are \emph{not} formulated as input-state-output \phss~which hampers their application to control design. \\
%
As can be seen from the above, the automated explicit port-Hamiltonian formulation of bond graphs has only been treated for special cases in literature so far. 
Ref.~\cite{lopes2016} and~\cite{donaire2009} address this topic but are restricted to particular classes of single-bond graphs.
Moreover, the literature lacks \emph{necessary} conditions for the existence of an explicit port-Hamiltonian formulation of bond graphs. 
The results of~\cite{lopes2016} suggest that an automated generation of port-Hamiltonian models from bond graphs is possible. 
However, a specific method which can be fully automated is missing. 
Lastly, all noted contributions on the port-Hamiltonian formulation of bond graphs focus on \emph{single}-bond graphs. To the best of our knowledge, a generalisation to \emph{multi}-bond graphs has not been addressed so far. 

\textbf{Contributions:}
This paper addresses the automated generation of explicit input-state-output \phss~from multi-bond graphs. 
The  main theoretical contributions are (i) the derivation of an explicit port-Hamiltonian formulation of multi-bond graphs and (ii) the presentation of two conditions, one necessary and one sufficient, for the existence of an explicit port-Hamiltonian formulation of multi-bond graphs. 
Furthermore, the main practical contribution of this paper is (iii) an algorithm which summarises the methods from (i) and (ii) in order to automatically generate an explicit \phs~from a given multi-bond graph.
Furthermore, we provide an implementation of (iii) in the Wolfram language (along with this publication).

\textbf{Paper organisation:}
In Section~\ref{sec:problem}, we define the problem under consideration. 
Section~\ref{sec:theomainresult} summarises the main theoretical result of this paper which is then interpreted and discussed. 
In Section~\ref{sec:bg2phs}, we provide the proof for the main theoretical result from Section~\ref{sec:theomainresult}. 
Section~\ref{sec:pracmainresult} assembles the results from Sections~\ref{sec:theomainresult} and~\ref{sec:bg2phs} in an overall algorithm which is the main practical result of this paper. 
Sections~\ref{sec:example} and \ref{sec:conclusion} provide an illustrative example and a conclusion of this paper, respectively. 

\textbf{Notation:}
Let $\vecspace{X}$ be a vector space.
For the dimension of $\vecspace{X}$ we write $\dim{\vecspace{X}}$.
Let $\vecspace{Y}$ be another vector space. 
$\vecspace{X}\isomorph\vecspace{Y}$ means that $\vecspace{X}$ and $\vecspace{Y}$ are isomorphic. \\
Let $\mat{A} = (a_{ij}) \in \R^{n\times m}$ be a matrix with $n$ rows and $m$ columns and $\x\in\R^{m}$ be a (column) vector. 
For a block diagonal matrix of matrices we write $\blkdiag(\cdot)$. 
We write $\mat{A}\posdef$ and $\mat{A}\possemidef$ if $\mat{A}$ is positive definite or positive semi-definite, respectively.
The image of the linear map \(\x \mapsto \mat{A}\x\) is written as \(\img \left(\mat{A}\right)\); 
for the kernel we write \(\ker \left( \mat{A} \right)\).
\(\set{O}(n)\) denotes the group of orthogonal matrices. 
The matrix $\0_{n\times m}$ is an $n\times m$ zero matrix;
we abbreviate $\0_{n\times n}$ to $\0_{n}$. 
Let $\blockZero_n^{p,q}$ denote an $(p\times q)$ block matrix of zero matrices $\0_n$. 
The $n\times n$ identity matrix is denoted as $\I_n$. 
$\blockI_n^{p \times q}$ is a $(p\times q)$ block matrix of identity matrices $\I_n$. \\
Let $\set{M}$ be a set.
$\card{\set{M}}$ denotes the cardinality of $\set{M}$.
For each $i\in\set{M}$, let $\mat{A}_i\in\R^{n\times m_i}$ be a matrix with $n$ rows and $m_i$ columns. 
For the \emph{horizontal} concatenation of all $\mat{A}_i$ we write $\left(\mat{A}_i\right)$ and append ``for all $i\in\set{M}$''. 
Further, for each $i\in\set{M}$, suppose a (column) vector $\x_i\in\R^{n}$. 
For the \emph{vertical} concatenation of all $\x_i$ we write $\left(\x_i\right)$ and append ``for all $i\in\set{M}$''. \\
Let $\mathcal{G} = \left(\V,\B\right)$ be a directed graph with vertices $\V$ and edges $\B$. 
The set of all adjacent vertices at $u\in \V$ is denoted as $\V(u) \define  \{v\in\V \mid (u,v)~\text{or}~(v,u) \in \B \}$. 
Suppose $\tilde{\B} \subseteq \B$. 
We define the set of all incident edges from $\tilde{\B}$ at $u\in \V$ as $\tilde{\B}(u) \define \{ (u,v),(v,u)\in \tilde{\B} \mid v\in \V \}$. 
Similarly, $\overleftarrow{\tilde{\B}}(u) \define \{ (v,u)\in \tilde{\B} \mid v\in \V \}$ and 
$\overrightarrow{\tilde{\B}}(u) \define \{ (u,v)\in \tilde{\B} \mid v\in \V \}$ are the sets of all ingoing and outgoing edges in $\tilde{\B}$ at $u\in \V$, respectively. 

\section{Problem definition} \label{sec:problem}
In this paper, we consider $\dimBond$-dimensional multi-bond graphs\footnote{In the remainder of this paper, we use ``bond graph'' for ``multi-bond graph'' and ``bond'' for ``multi-bond''.} ($\dimBond \in \N_{\geq 1}$) in the generalised bond graph framework \cite[p. 24]{duindam2009} with   
the following types of elements: storages~(\bgC), modulated resistors~(\bgR), sources of flow~(\bgSf), sources of effort~(\bgSe), 0-junctions~(\bgZero), 1-junctions~(\bgOne), modulated transformers~(\bgTF) and modulated gyrators~(\bgGY). 
%
Let the set $\set{E} \define \{ \bgC, \bgR, \bgSf, \bgSe, \bgZero, \bgOne, \bgTF, \bgGY \}$ collect the different types of elements. \\
%
In the sequel, we describe the topology of a bond graph by a directed graph. 
For each $\alpha\in\set{E}$, let us define a set $\V_\alpha$ with $\numV_\alpha \define \card{\V_\alpha}$ which contains all elements of type $\alpha$.
We will denote elements of type $\bgC$, $\bgR$, $\bgSf$, $\bgSe$ as \emph{exterior elements}; elements of type \bgZero, \bgOne, \bgTF, \bgGY~are referred to as \emph{interior elements}. The sets of exterior and interior elements are defined as
\sca{possible space save: inline formula}
\begin{subequations} \label{eq:exterior_interior:vertices}
\begin{align} 
\V_\bgE & \define \V_\bgC \cup \V_\bgR \cup\V_\bgSf\cup \V_\bgSe, \label{eq:vertices:external}\\
\V_\bgI & \define \V_\bgZero \cup  \V_\bgOne \cup \V_\bgTF \cup\V_\bgGY, \label{eq:vertices:internal}
\end{align}
\end{subequations}
with $\numV_\bgE \define \card{\V_\bgE}$ and $~\numV_\bgI \define \card{\V_\bgI}$, respectively. The union $\V \define \cup_{\alpha \in \set{E}} \V_\alpha = \cup_{\alpha \in \{\bgE,\bgI\}} \V_\alpha$ is the set of all bond graph elements ($\numV \define \card{\V}$).  
The $\numV$ elements of $\V$ are connected by a set $\set{B}$ of $\numB$ bonds, i.e. $\numB\define\card{\set{B}}$. Each bond $j\in\set{B}$ carries a flow $\f_j \in \R^\dimBond$ and an effort $\e_j \in \R^\dimBond$. The directed graph $\mathcal{G}=\left(\V,\B\right)$ describes the topology of the bond graph. Analogous to the naming of elements, we define sets of exterior and interior bonds
\begin{subequations} \label{eq:exterior_interior:edges}
\begin{align} 
\B_\bgE & \define \{ (u,v),(v,u) \in\B \mid v\in \V_\bgE, u \in \V_\bgI \},\\
\B_\bgI & \define \{ (u,v) \in\B \mid u,v\in \V_\bgI \},
\end{align}
\end{subequations}
with $\numB_\bgE \define \card{\B_\bgE}$ and $\numB_\bgI \define \card{\B_\bgI}$. The set
$\B_\bgE$ contains bonds which connect an exterior element to an interior element; $\B_\bgI$ contains bonds which connect two interior elements with each other.  \\
We consider bond graphs that are non-degenerate, i.e. bond graphs where $\mathcal{G}=\left(\V,\set{B}\right)$ is weakly connected, and where each exterior element is connected by exactly one bond to one interior element, i.e.\ for each $v \in \V_\bgE$ we have $\V(v) \subset \V_\bgI$ with $\card{\B(v)}=1$ and $\card{\V(v)}=1$. 
Moreover, we use the bond orientation rules from standard bond graph literature \cite[p.~59]{borutzky2010} in which bonds are incoming to storages and resistors and outgoing from sources of flow and effort. 
Without loss of generality, we assume each transformer and each gyrator to have exactly one incoming and exactly one outgoing bond in order to enable an unambiguous definition of transformer and gyrator ratios. 
\begin{defn} \label{def:js}
	The \emph{junction structure} of a bond graph is defined as the sub-graph $\mathcal{G}_\bgI \subset \mathcal{G}$ with $\mathcal{G}_\bgI=\left(\V_\bgI,\B_\bgI\right)$. 
\end{defn}
From the properties of a non-degenerate bond graph, it follows that $\mathcal{G}_\bgI$ is weakly connected and $\B = \B_\bgE \cup \B_\bgI$. 
Furthermore, we make the following two assumptions. 
\begin{assum}\label{ass:modulation}
Modulation of resistors, transformers and gyrators can be expressed only in dependence on states of $\bgC$-type elements and constant parameters.
\end{assum}
\begin{assum} \label{ass:linearResistors}
The constitutive relations of modulated resistors are linear with respect to the respective power-port variables and in Onsager form~\cite[p.~364]{borutzky2010}. 
\end{assum}
In \cite[p. 159]{borutzky2010}, it is shown that bond graphs violating Assumption~\ref{ass:modulation} \emph{cannot} in general be formulated in an explicit form.  
Likewise, Assumption \ref{ass:linearResistors} is a well-known requirement for formulating an explicit \phs~\cite[p.~53]{van_der_schaft2014}. 
Next, we define the mathematical representation of interest in this paper.  
\begin{defn} \label{def:phs}
An \emph{explicit input-state-output port-Hamiltonian system (\phs)} (with feedthrough) is defined as dynamic system of the form
\begin{subequations}
\label{eq:phs}
\begin{align}
	\label{eq:dynphs}
	\dot{\x} &= \left[ \matJ\ofx -\matR \ofx \right] \tfrac{\partial H}{\partial \x} \ofx + \left[\matG\ofx-\mat{P}\ofx \right] \vecu, \\
	\label{eq:outputphs}
	\vecy 	&= \left[ \matG\ofx + \mat{P} \ofx\right]^{\T} \! \tfrac{\partial H}{\partial \x} \ofx + \left[\matM\ofx + \matS\ofx \right] \vecu,
\end{align}
\end{subequations}
where $\x \in \X$, $\vecu \in \R^p$, and $\vecy \in \R^p$ are the \emph{state vector}, the \emph{input vector}, and the \emph{output vector}, respectively\footnote{Throughout this paper we omit the time-dependence ``$(t)$'' of vectors in the notation.}. 
We assume the state-space $\X$ to be a real vector space with $\dim\X=n$. 
The Hamiltonian is a non-negative function $H \colon \X \to \R_{\geq 0}$. 
The matrices $\matJ\ofx$, $\matR\ofx \in \R^{n\times n}$, $\matG\ofx$, $\matP\ofx \in \R^{n\times p}$, $\matM\ofx$, $\matS\ofx \in \R^{p\times p}$ satisfy $\matJ\ofx=-\matJ\Tofx$, $\matM\ofx=-\matM\Tofx$, and
\begin{equation} \label{eq:phs:definiteness}
	\mat{Q}\ofx \define
	\begin{pmatrix}
	\matR\ofx & \mat{P}\ofx \\
	\mat{P}\Tofx & \matS \ofx
	\end{pmatrix}  = \mat{Q}\Tofx
	\possemidef,~ \forall \x\in\X.
\end{equation}
\end{defn}
With the following property, we exclude causally implausible port-Hamiltonian formulations of bond graphs. To this end, we require the flows of $\bgSf$ elements and the efforts of $\bgSe$ elements to act as inputs in the \phs. Correspondingly, the respective conjugated variables must act as output of the \phs. 
\begin{req}\label{req:inputs_outputs}
Let $\B_\alpha = \cup_{i\in\V_\alpha} \B(i)$ for $\alpha\in\{\bgSf,\bgSe\}$. In \eqref{eq:phs}, $\vecu$ consists of $(\f_j)$, $(\e_k)$  while the $\vecy$ consists of $(\e_j)$, $(\f_k)$ for all $j\in\B_\bgSf$, $k\in\B_\bgSe$.
\end{req}
This paper will show a solution to the following problem:
\begin{prob}
Consider an $\dimBond$-dimensional bond graph that satisfies Assumptions~\ref{ass:modulation} and~\ref{ass:linearResistors}. 
What is a constructive and automatable method that formulates the bond graph as a \phs~\eqref{eq:phs} with Property~\ref{req:inputs_outputs}.
\end{prob}

\section{Main theoretical result} \label{sec:theomainresult}
In this section, we present and discuss the main theoretical result of this paper. 
This main result is summarised in Theorem~\ref{theorem:main_result} which contains a structured method to formulate an $\dimBond$-dimensional bond graph as an explicit \phs~with Property~\ref{req:inputs_outputs}. 
The theorem is organised in four parts: 
In (i), the junction structure of the bond graph is described by a Dirac structure in implicit representation\footnote{cf. Remark~\ref{rem:DS:imex}.}. 
Afterwards, in (ii) the Dirac structure is transferred from an implicit to an explicit representation. 
The inputs and outputs in the explicit representation are chosen under consideration of Property~\ref{req:inputs_outputs}. 
In (iii), the explicit representation of the Dirac structure is merged with the constitutive relations of storages and resistors which leads to an explicit port-Hamiltonian formulation of the bond graph. 
Finally, part (iv) provides two conditions, one necessary and one sufficient, for the existence of such an explicit formulation.  
A discussion of Theorem~\ref{theorem:main_result} concludes this section. 
Preliminaries on Dirac structures are given in Appendix~\ref{sec:preliminaries}. 
\begin{thm} \label{theorem:main_result}
(i) Given an $\dimBond$-dimensional bond graph that satisfies Assumption~\ref{ass:modulation}, the junction structure of the bond graph can be described by a Dirac structure in implicit form:
\begin{multline}
\label{eq:ds:kernel:blocks}
\dirac = \{ 
(\begin{pmatrix}
	\f_\bgC \\	
	\f_\bgR \\
	\f_\bgSf \\
	\f_\bgSe \\
	\end{pmatrix},
	\begin{pmatrix}
	\e_\bgC \\	
	\e_\bgR \\
	\e_\bgSf \\
	\e_\bgSe \\
	\end{pmatrix} )
	\in \R^{\dimBond\numB_\bgE} \times \R^{\dimBond\numB_\bgE}
	\mid  \\ 
	\underbrace{
	\begin{pmatrix} \F_\bgC\Tofx \\ \F_\bgR\Tofx \\ \F_\bgSf\Tofx \\ \F_\bgSe\Tofx  \end{pmatrix}^\T}_{\definer\F\ofx}
	\begin{pmatrix}
	-\f_\bgC \\
	-\f_\bgR \\
	\f_\bgSf \\
	\f_\bgSe \\
	\end{pmatrix}
	+  
	\underbrace{
	 \begin{pmatrix} \E_\bgC\Tofx \\ \E_\bgR\Tofx  \\ \E_\bgSf\Tofx \\ \E_\bgSe\Tofx \end{pmatrix}^\T}_{\definer \E\ofx}
	\begin{pmatrix}
	\e_\bgC \\
	\e_\bgR \\
	\e_\bgSf \\
	\e_\bgSe \\
	\end{pmatrix}
	= \0
	\}.
\end{multline}
where $\f_\alpha= \left(\f_i\right)\in\R^{\dimBond\numV_\alpha}$, $\e_\alpha = \left(\e_i\right) \in\R^{\dimBond\numV_\alpha}$ for all $i \in \V_\alpha$ and $\F_\alpha \ofx ,\E_\alpha \ofx \in\R^{\dimBond\numV_\bgE \times \dimBond\numV_\alpha}$ with $\alpha\in\{\bgC,\bgR,\bgSf,\bgSe\}$.\footnote{The negative sign of $\f_\bgC$ and $\f_\bgR$ in \eqref{eq:ds:kernel:blocks} stems from the fact that bonds are \emph{incoming} to storages and resistors.} 

(ii) Let the matrices in \eqref{eq:ds:kernel:blocks} fulfill 
\begin{equation} \label{eq:FEF:fullrank}
\rank\left(\F_\bgC\ofx ~ \E_\bgSf \ofx  ~ \F_\bgSe \ofx  \right) = \dimBond \left( \numV_\bgC + \numV_\bgSf + \numV_\bgSe\right),
\end{equation}
for all $\x\in\X$. Then, \eqref{eq:ds:kernel:blocks} can be formulated in an explicit representation
\begin{multline}
\label{eq:dirac:io:specialform}
\dirac = \{ 
(
\begin{pmatrix}
	\f_\bgC \\
	\f_\bgR \\
	\f_\bgSf \\
	\f_\bgSe 
	\end{pmatrix},
	\begin{pmatrix}
	\e_\bgC \\	
	\e_\bgR \\
	\e_\bgSf \\
	\e_\bgSe 
	\end{pmatrix}
)
	\in \R^{\dimBond\numV_\bgE} \times \R^{\dimBond\numV_\bgE}
	\mid \\
	\begin{pmatrix}
	\vecy_\bgC \\
	\vecy_\bgR \\
	\vecy_\bgP 
	\end{pmatrix} = 
\underbrace{\begin{pmatrix*}[r]
	\matZ_{\bgC\bgC}\ofx 	& -\matZ_{\bgC\bgR} \ofx 	& -\matZ_{\bgC\bgP}\ofx \\
	 \matZ_{\bgC\bgR}^\T\ofx & \matZ_{\bgR\bgR}\ofx  	& -\matZ_{\bgR\bgP}\ofx \\
	 \matZ_{\bgC\bgP}^\T\ofx	& \matZ_{\bgR\bgP}^\T \ofx 	& \matZ_{\bgP\bgP}\ofx 
	\end{pmatrix*}}_{\matZ\ofx}
	\begin{pmatrix}
	\vecu_\bgC \\	
	\vecu_\bgR \\
	\vecu_\bgP 
	\end{pmatrix}
	\},
\end{multline}
where $\matZ\ofx=-\matZ^\T\ofx$ for all $\x\in\X$ with
\begin{subequations} \label{eq:iods:Zuy}
\begin{align} 
		\mat{Z}\ofx &=
		\left( \F_{\bgC}\ofx ~ \F_{\bgR,1}\ofx ~ \E_{\bgR,2}\ofx ~ \E_\bgSf\ofx ~ \F_\bgSe\ofx \right)^{-1} \notag \\
		& \quad \cdot  \left( \E_{\bgC}\ofx ~ \E_{\bgR,1}\ofx ~ \F_{\bgR,2}\ofx ~ \F_\bgSf\ofx ~ \E_\bgSe\ofx \right) \label{eq:formula:Z_theorem}
\shortintertext{and}
	\vecu_\bgC &= \e_{\bgC}, 	\label{eq:dirac:io:u}
	~~~ \vecu_\bgR = \begin{pmatrix*}[r] \e_{\bgR,1} \\  -\! \f_{\bgR,2}	\end{pmatrix*}, 	
	~~ \vecu_\bgP = \begin{pmatrix} \f_\bgSf	\\ \e_\bgSe \end{pmatrix}, \\
	\vecy_\bgC &= - \f_{\bgC},	\label{eq:dirac:io:y}
	~ \vecy_\bgR = \begin{pmatrix*}[r] - \f_{\bgR,1} \\ \e_{\bgR,2}	\end{pmatrix*}, 
	~\, \vecy_\bgP = \begin{pmatrix} \e_\bgSf \\ \f_\bgSe \end{pmatrix}. 
\end{align}
\end{subequations}
In \eqref{eq:formula:Z_theorem}, $\left( \F_{\bgR,1}\ofx\, \F_{\bgR,2}\ofx \right)$ is a splitting of $\F_{\bgR} \ofx$ (possibly after some permutations)  such that (a) $\left(\F_{\bgC}\ofx \, \F_{\bgR,1}\ofx \, \E_\bgSf\ofx \, \F_\bgSe\ofx \right)$ has full column rank and (b) $\rank \left(\F_{\bgC}\ofx ~ \F_{\bgR,1}\ofx ~ \E_\bgSf\ofx ~ \F_\bgSe\ofx \right)$ is equal to $\rank \left( \F_{\bgC}\ofx ~ \F_{\bgR}\ofx ~ \E_\bgSf\ofx ~ \F_\bgSe\ofx \right)$ for all $\x\in\X$. Such a splitting of $\F_{\bgR} \ofx$ always exists. According to the splitting of $\F_{\bgR} \ofx$, we split $\E_{\bgR} \ofx$ into $\left(\E_{\bgR,1}\ofx \, \E_{\bgR,2}\ofx \right)$ and the vectors $\f_{\bgR}$ and $\e_{\bgR}$ (see $\vecu_\bgR$ and $\vecy_\bgR$ in \eqref{eq:dirac:io:u} and \eqref{eq:dirac:io:y}, respectively). 

(iii) For the bond graph, suppose $\bgC$-type elements subject to nonlinear constitutive relations of the form~\cite[pp.~357--358]{borutzky2010}
\begin{align} \label{eq:storagePort}
	\vecy_\bgC = -\f_\bgC = -\vec{\dot{x}}, \quad\quad \vecu_\bgC=\e_\bgC=\frac{\partial H}{\partial \x} \ofx. 
\end{align}
with energy state $\x\in\X$, $\dim(\X)=\dimBond\numVC$, and energy storage function $H: \X \to \R_{\geq 0}$, $\x \mapsto H(\x)$. 
Further, let Assumption~\ref{ass:linearResistors} hold, which enables us to write the constitutive relations of the $\bgR$-type elements as
\begin{equation} \label{eq:resistivePort}
	\f_\bgR= \mat{D} \ofx \e_\bgR
\end{equation}
where $\mat{D} \ofx = \mat{D} \ofx ^\T \possemidef$. 
Assume that \eqref{eq:resistivePort} can be written in input-output form
\begin{equation} \label{eq:ass:R-port:theo}
	\vecu_\bgR = -\tilde{\mat{R}} \ofx \vecy_\bgR
\end{equation}
with $\tilde{\mat{R}}\ofx=\tilde{\mat{R}}\ofx^\T \possemidef$.
The bond graph can then be formulated as explicit input-state-output \phs~of the form \eqref{eq:phs} with state $\x$ and Hamiltonian $H\ofx$ from \eqref{eq:storagePort}. Moreover, the inputs and outputs of the \phs~are given by $\vecu = \vecu_\bgP$, $\vecy = \vecy_\bgP$ from \eqref{eq:dirac:io:u} and \eqref{eq:dirac:io:y}, respectively. Thus, the \phs~has Property~\ref{req:inputs_outputs}. 
The matrices of \eqref{eq:phs} are calculated as:
\begin{subequations}
\label{eq:real:phs_matrices_from_HIO_DS}
\begin{align}
	\matJ\ofx &\!= -\matZ_{\bgC\bgC}\ofx-\tfrac{1}{2}\matZ_{\bgC\bgR}\ofx \mat{A}\ofx\matZ_{\bgC\bgR}^\T\ofx, \\
	\matR\ofx	&\!= \tfrac{1}{2}\matZ_{\bgC\bgR}\ofx \mat{B}\ofx\matZ_{\bgC\bgR}^\T\ofx, \\
	\matG\ofx &\!= \matZ_{\bgC\bgP}\ofx + \tfrac{1}{2}\matZ_{\bgC\bgR}\ofx \mat{A}\ofx\matZ_{\bgR\bgP}\ofx, \\
	\matP\ofx	&\!= -\tfrac{1}{2}\matZ_{\bgC\bgR}\ofx \mat{B}\ofx\matZ_{\bgR\bgP}\ofx, \\
	\matM\ofx &\!= \matZ_{\bgP\bgP}\ofx+\tfrac{1}{2}\matZ_{\bgR\bgP}^\T\ofx \mat{A}\ofx\matZ_{\bgR\bgP}\ofx, \\
	\matS\ofx	&\!= \tfrac{1}{2}\matZ_{\bgR\bgP}^\T \ofx \mat{B}\ofx\matZ_{\bgR\bgP}\ofx,
\shortintertext{where}
	\mat{A}\ofx &\!= \tilde{\mat{K}}\ofx\tilde{\matR}\ofx-\tilde{\matR}\ofx\tilde{\mat{K}}^\T\ofx,\label{eq:iods2phs:A} \\
	\mat{B}\ofx &\!= \tilde{\mat{K}}\ofx\tilde{\matR}\ofx+\tilde{\matR}\ofx\tilde{\mat{K}}^\T \ofx, \label{eq:iods2phs:B} \\
	\tilde{\mat{K}}\ofx &\!= (\I+\tilde{\matR}\ofx\matZ_{\bgR\bgR}\ofx)^{-1}. \label{eq:Ktilde}
\end{align}
\end{subequations}
(iv) Equations~\eqref{eq:FEF:fullrank} and~\eqref{eq:ass:R-port:theo} together form a \emph{sufficient} condition for the existence of an explicit formulation~\eqref{eq:phs} of a bond graph.  
Moreover, \eqref{eq:FEF:fullrank} implies 
\begin{equation} \label{eq:theo:EF:fullrank}
\rank\left(\E_\bgSf \ofx  ~ \F_\bgSe \ofx  \right) = \dimBond \left( \numV_\bgSf + \numV_\bgSe\right),~\forall\x\in\X.
\end{equation}
which is (under Property~\ref{req:inputs_outputs}) a \emph{necessary} condition for the existence of such a formulation. 
\end{thm}
Theorem~\ref{theorem:main_result} gives a structured method to formulate a bond graph as \phs~\eqref{eq:phs} with Property~\ref{req:inputs_outputs}. 
The matrices of the \phs~can be calculated with the equations in \eqref{eq:real:phs_matrices_from_HIO_DS}. 
These equations reveal that the matrices of the \phs~are \emph{independent} of the storage function of the $\bgC$-type elements in \eqref{eq:storagePort}. 
Conversely in \eqref{eq:storagePort}, the state vector and the Hamiltonian of the \phs~are solely dependent on variables and parameters of $\bgC$-type elements.
Hence, the separation of energy-storage elements and energy-routing elements of the bond graph directly translates into the explicit \phs. 
%
By \eqref{eq:ds:kernel:blocks}, \eqref{eq:dirac:io:specialform}, and \eqref{eq:real:phs_matrices_from_HIO_DS}, we see that state-modulated transformers and gyrators yield an explicit \phs~with state-dependent matrices. 
Similarly, state-modulated $\bgR$-type elements generally results in a state-dependent \phs~matrices. 
If all bond graph elements of type $\bgTF$, $\bgGY$, and $\bgR$ are \emph{unmodulated}, the matrices of the explicit port-Hamiltonian formulation in \eqref{eq:real:phs_matrices_from_HIO_DS} are constant. 
If, in addition, the storages obey quadratic storage functions, the resulting \phs~is linear. 
In conclusion, the major properties of a bond graph translate into the explicit port-Hamiltonian formulation. 
Thus, \eqref{eq:phs} may be seen as a natural explicit state-space representation of bond graphs. 
\begin{rem} \label{rem:Ktilde}
In Lemma~\ref{lemma:regularityK} we will show that the matrix $\tilde{\mat{K}}\ofx$ in \eqref{eq:Ktilde} always exists. 
This matrix (or related expressions) has appeared in previous publications addressing the derivation of state-space formulations of bond graphs, e.g. \cite[eq. (7)]{rosenberg1971}, \cite[eq. (29)]{wellstead1979}, \cite[eq. (14)]{donaire2009}, and \cite[Remark~2]{lopes2016}. However, to the best of the authors' knowledge, the existence of $\tilde{\mat{K}}$ has not been discussed so far. \\
\end{rem}
Equation~\eqref{eq:theo:EF:fullrank} is a necessary condition for the existence of an input-state-output model that has Property~\ref{req:inputs_outputs}. 
This condition is plausible as it prevents the bond graph from having \emph{dependent sources}~\cite{borutzky2010} which are physically implausible~\cite[p.~169]{karnopp2012}.  
\begin{rem} \label{rem:implicit:phs}
Equation~\eqref{eq:theo:EF:fullrank} is also necessary if we aim at an \emph{implicit} port-Hamiltonian formulation of a bond graph~\cite{golo2003} which has Property~\ref{req:inputs_outputs}. 
This is plausible as~\eqref{eq:theo:EF:fullrank} is necessary for a bond graph to be \emph{well-posed} in the sense of~\cite[Def.~2]{golo2003}.  
\end{rem}
Together, \eqref{eq:FEF:fullrank} and~\eqref{eq:ass:R-port:theo} form a sufficient condition for the existence of an explicit port-Hamiltonian formulation of a bond graph. 
Equation~\eqref{eq:FEF:fullrank} is more stringent than \eqref{eq:theo:EF:fullrank} as it, in addition to dependent sources, prevents the bond graph from having 
(i) dependent storages~\cite[p.~107]{borutzky2010} and 
(ii) storages that are directly determined by source elements. 
From bond graph theory, it is known that (i) and (ii) occur from physically implausible structures in the bond graph. 
Moreover, different strategies exist to resolve such implausible structures in the bond graph \cite{borutzky2010}.
Thus, \eqref{eq:FEF:fullrank} is not very restrictive. 
Equation~\eqref{eq:ass:R-port:theo} assumes the resistive structure to be in an input-output form which is a well-known requirement for the derivation of explicit input-state-output \phss~\cite[p.~53]{van_der_schaft2014}. 
For many bond graphs, \eqref{eq:ass:R-port:theo} is satisfied by design. In particular, for single-bond graphs ($\dimBond=1$) equation~\eqref{eq:ass:R-port:theo} is always fulfilled. 
\begin{rem} \label{rem:DAE}
Theorem~\ref{theorem:main_result} is independent of the particular form of the skew-symmetric matrix $\matZ_{\bgR\bgR} \ofx$ in \eqref{eq:dirac:io:specialform}. 
This is remarkable as dependent resistors (i.e. $\matZ_{\bgR\bgR} \ofx \neq \0$) are generally known to lead to models in the form of differential-algebraic equations \cite[p.~134]{borutzky2010}, \cite[p.~187]{karnopp2012}. 
\end{rem}

\section{Proof of Theorem~\ref{theorem:main_result}} \label{sec:bg2phs}
In this section, we present a constructive proof of Theorem~\ref{theorem:main_result}. As the theorem, the proof is subdivided into four parts. Each of the following Sections~\ref{subsec:JS2DSs} to \ref{sec:proof:thm2} is dedicated to the corresponding part (i) to (iv) of Theorem~\ref{theorem:main_result}. 

%
\subsection{Description of interior elements as Dirac structures} \label{subsec:JS2DSs}
In this section, we show that the junction structure of the bond graph can always be described by a Dirac structure of the form~\eqref{eq:ds:kernel:blocks}.
The approach is as follows: 
First, we show that the constitutive relations of the set of interior elements of a bond graph can be described as a set of Dirac structures for which we provide specific matrix representations. 
Secondly, we present an approach to compose the set of Dirac structures to one single Dirac structure.
Preliminaries on Dirac structures are given in Appendix~\ref{sec:preliminaries}. \\ 
Before we formulate specific Dirac structures for the interior elements, we give two preliminary statements. 
\begin{lem} \label{lemma:orthogonal:matrix:DS}
Suppose a modulated Dirac structure~\eqref{eq:ds:kernel} and let \(\mat{T}\ofx \in \set{O}(n)\) be a family of orthogonal matrices parametrised over $\x\in\X$. Then
\begin{align} \label{eq:ds:kernel:permutation}
	\tilde{\dirac}\ofx \tighteq \lbrace
	(\tilde{\f} ,\tilde{\e} ) \in \R^n \! \times \! \R^n
	\! \mid \! \tilde{\F}\ofx \! \tilde{\f}+ \tilde{\E}\ofx\!\tilde{\e}= \0 \rbrace
\end{align}
with $\tilde{\F}\ofx \! \tighteq\!  \F\ofx \! \mat{T}\ofx^\T$\!\!, $\tilde{\E}\ofx \! \tighteq \! \E\ofx \! \mat{T}\ofx^\T$\! is a modulated Dirac structure. 
\end{lem}
\begin{pf}
Inserting $\f = \mat{T}\ofx^\T \tilde{\f}$ and $\e = \mat{T}\ofx^\T \tilde{\e}$ into \eqref{eq:ds:kernel} gives \eqref{eq:ds:kernel:permutation}. Equation~\eqref{eq:ds:kernel:permutation} is a Dirac structure as it fulfills~\eqref{eq:dirac:kernel:cond}:
\begin{subequations}
\label{eq:dirac:kernel:permute:cond}
		\begin{alignat}{2}
		&(i)~&&\tilde{\F}\ofx \tilde{\E}\Tofx  + \tilde{\E}\ofx \tilde{\F}\Tofx = \notag \\
		&~&&\qquad \qquad \E \ofx \F^\T \ofx +\F\ofx \E\Tofx =\0,    \\
		&(ii)~&&\rank (\tilde{\F}\ofx~\tilde{\E}\ofx)=\rank\left( (\F\ofx~\E\ofx)\mat{T}\ofx{}^{\!\T}  \right) = \notag \\
		&~&&\qquad \qquad \rank (\F\ofx~\E\ofx) = n. \qed
		\end{alignat}
\end{subequations}
\end{pf}
%
%
\begin{cor} \label{cor:equivalentDS}
Given two vector spaces
\begin{equation}
				\dirac_i\ofx =	\lbrace	(\f_i,\e_i)  \in \R^n \times \R^n   \mid    \F_i\ofx  \f_i + \E_i\ofx  \e_i = \vec{0}\rbrace
\end{equation}
with $\x\in\X$, $i\in\{1,2\}$. If for every \(\x\in\X\) there exists a ~$\mat{T}\ofx\in \set{O}(n)$ such that $(\f_1,\e_1) \mapsto (\mat{T}\ofx\f_1,\mat{T}\ofx\e_1)$ is a bijection between $\dirac_1\ofx$ and $\dirac_2\ofx$, then
``$\dirac_1\ofx$ is a Dirac structure'' is equivalent to ``$\dirac_2\ofx$ is a Dirac structure''.
%
\end{cor}
\begin{pf}
The proof follows directly from a twofold application of Lemma~\ref{lemma:orthogonal:matrix:DS}.
\end{pf}
The following lemma now provides specific matrix representations of Dirac structures describing the constitutive relations of each interior element. 
\begin{lem} \label{lemma:DS:patterns}
Given an $\dimBond$-dimensional bond graph which fulfills Assumption~\ref{ass:modulation}. 
Let us consider the set of interior elements $\V_\bgI$ from \eqref{eq:vertices:internal} with $\numV_\bgI = \card{\V_\bgI}$. 
The constitutive relations of all elements of $\V_\bgI$ can be described by a set of Dirac structures $\set{DS}$ with $\card{\set{DS}} = \numV_\bgI$. For each element $i \in \V_\bgI$ there exists a corresponding Dirac structure $\dirac_i\ofx \in \set{DS}$ with
\begin{multline} \label{eq:KDSs}
\dirac_i\ofx = \{ 
 ( \begin{pmatrix}
	\left(\f_j\right)\\
	\left(\f_k\right)
	\end{pmatrix},
	\begin{pmatrix}
	\left(\e_j\right) \\
	\left(\e_k\right)
	\end{pmatrix} ) \in \R^{\dimBond \cdot \numB(i)} \times \R^{\dimBond \cdot \numB(i)} \mid \\
 \F_i\ofx
\begin{pmatrix}
	\left(\f_j\right)\\
	-\left(\f_k\right)
	\end{pmatrix} + 
\E_i\ofx
	\begin{pmatrix}
	\left(\e_j\right) \\
	\left(\e_k\right)
\end{pmatrix}
	= \0
	\},
\end{multline}
for all $j\in\Bin(i),k\in\Bout(i)$, and $\numB(i) \define \card{\B(i)}$.  
Depending on the type of $i$, the matrices $\F_i\ofx$ and $\E_i\ofx$ in \eqref{eq:KDSs} are as follows. 
For $i\in\V_\bgZero$ and $i\in\V_\bgOne$ we have
\begin{subequations}\label{eq:patterns:01}
\begin{align}
&\F_i = \mat{\Psi}_i, &\E_i = \mat{\Theta}_i; \label{eq:FE:0} \\
\shortintertext{and}
&\F_i = \mat{\Theta}_i\mat{T}_i, &\E_i = \mat{\Psi}_i\mat{T}_i; \label{eq:FE:1} 
\end{align}
\end{subequations}
respectively, 
\begin{subequations} \label{eq:psitheta}
\begin{align} 
\mat{\Psi}_i &=  \begin{pmatrix} \blockI_\dimBond^{1\times \numB(i)} \\  \blockZero_\dimBond^{(\numB(i)-1)\times \numB(i)} \end{pmatrix}, \\
\mat{\Theta}_i &= \begin{pmatrix} \0_N & \blockZero_N^{1\times(\numB(i)-1)} \\  \blockI_\dimBond^{(\numB(i)-1)\times1} & -\I_{N(\numB(i)-1)} \end{pmatrix},
\end{align}
\end{subequations} 
and $\mat{T}_i=\blkdiag\left( \I_{\dimBond \cdot \card{\Bin(i)}}, -\I_{\dimBond \cdot \card{\Bout(i)}} \right)$. For $i\in\V_\bgTF$ and $i\in\V_\bgGY$ the matrices are given by
\begin{subequations}\label{eq:patterns:TFGY}
\begin{align}
&\F_i\ofx  = \begin{pmatrix} \I_\dimBond & \mat{U}_i \ofx\\  \0_\dimBond & \0_\dimBond \end{pmatrix} \!,
~\E_i\ofx  = \begin{pmatrix} \0_\dimBond & \0_\dimBond \\  -\mat{U}^\T_i\ofx & \I_\dimBond \end{pmatrix}\!; \label{eq:FE:TF} \\
\shortintertext{and}
&\F_i\ofx  = \begin{pmatrix} \0_\dimBond & \mat{V}_i\ofx\\  -\mat{V}_i^\T\ofx & \0_\dimBond \end{pmatrix}\!, 
~\E_i  =  \begin{pmatrix} \I_\dimBond & \0_\dimBond\\  \0_\dimBond & \I_\dimBond \end{pmatrix}\!, \label{eq:FE:GY}
\end{align}
\end{subequations}
where $\mat{U}_i\ofx$ and $\mat{V}_i\ofx$ are square matrices of full rank $\dimBond$ for all $\x\in\X$, which describe the (multi-dimensional) transformer and gyrator ratios, respectively.  

\end{lem}
\begin{pf}
First, we prove that each element $\dirac_i\ofx\in\set{DS}$ describes the constitutive relations of the corresponding interior element $i\in\V_\bgI$. Secondly, we show that the elements $\dirac_i\ofx\in\set{DS}$ define Dirac structures. \\
For $i\in\V_\bgZero$, we insert the matrices \eqref{eq:FE:0} with \eqref{eq:psitheta} into the equation system of \eqref{eq:KDSs} and obtain Kirchhoff's current law which is the relation governing $\bgZero$-junctions.  
Analogously, for $i\in\V_\bgOne$ we obtain Kirchhoff's voltage law in the form $\tilde{\F_i}\tilde{\f_i}+\tilde{\E_i}\tilde{\e_i}=\0$ with 
\begin{equation} \label{eq:KDS:1:aux1}
\mat{\Theta}_i
\begin{pmatrix}
	\left(\tilde{\f}_j\right)\\
	\left(\tilde{\f}_k\right)
\end{pmatrix} + 
\mat{\Psi}_i
\begin{pmatrix}
	\left(\tilde{\e}_j\right)\\
	-\left(\tilde{\e}_k \right)
\end{pmatrix} = \0, 
\end{equation}
for all $j\in\Bin(i),k\in\Bout(i)$. To bring \eqref{eq:KDS:1:aux1} to the form of the equation system in \eqref{eq:KDSs}, we perform a change of coordinates ${\f_i} = \mat{T}_i^\T \tilde{\f_i}$, ${\e_i} = \mat{T}_i^\T \tilde{\e_i}$, with matrix $\mat{T}_i$ as above to obtain \eqref{eq:FE:1}. 
For $i \in \V_\bgTF$ and $i \in \V_\bgGY$ we insert \eqref{eq:FE:TF} and \eqref{eq:FE:GY} into the equation system of \eqref{eq:KDSs} and get 
\begin{align}	
& i\in\V_\bgTF: &\f_j = \mat{U}_i\ofx \f_k,\quad & \e_k= \mat{U}^\T_i\ofx\e_j, \label{eq:constEq:TF} \\
& i\in\V_\bgGY: &\e_j = \mat{V}_i\ofx \f_k,\quad & \e_k = \mat{V}^\T_i\ofx\f_j, \label{eq:constEq:GY} 
\end{align}
where $j \in \Bin(i), k \in \Bout(i)$. Equations~\eqref{eq:constEq:TF} and~\eqref{eq:constEq:GY} are the relations governing multi-dimensional transformers and gyrators, respectively \cite[pp.~358--359]{borutzky2010}. 
Inserting the matrices $\F_i\ofx$ and $\E_i\ofx$ from \eqref{eq:FE:0}, \eqref{eq:FE:TF}, \eqref{eq:FE:GY} into \eqref{eq:dirac:kernel:cond} shows that these matrices indeed define Dirac structures. Analogously, the matrices from \eqref{eq:KDS:1:aux1} define a Dirac structure. As $\mat{T}_i\in\set{O}(\numB(i))$, by Corollary \ref{cor:equivalentDS} the matrices~\eqref{eq:FE:1} then also define a Dirac structure. 
\qed \end{pf}
%
%
%
%
Lemma~\ref{lemma:DS:patterns} provides a set $\set{DS}$ containing $\numV_\bgI$ Dirac structures. The $\numV_\bgI$ Dirac structures describe the constitutive equations of the $\numV_\bgI$ interior elements of the bond graph by relating the flows and efforts of the exterior and interior bonds. 
In the sequel, we show that it is always possible to compose the $\numV_\bgI$ Dirac structures to one \emph{single} Dirac structure~\eqref{eq:ds:kernel:blocks} which relates the flows and efforts of only the exterior bonds, i.e. \emph{without} using flows and efforts of interior bonds. 
For the composition, we use the methods from \cite{batlle2011} and \cite[pp.~70ff.]{van_der_schaft2014}. \\
Consider the sets of exterior and interior vertices $\V_\bgE$, $\V_\bgI$ and the sets of exterior and interior bonds $\B_\bgE$, $\B_\bgI$ as defined in~\eqref{eq:exterior_interior:vertices} and~\eqref{eq:exterior_interior:edges}, respectively. 
From $\B=\B_\bgE \cup \B_\bgI$ it follows that for each $i\in\V_\bgI$ we can reorder (cf.\ Corollary~\ref{cor:equivalentDS}) the vectors and matrices of $\dirac_i\in\set{DS}$ in~\eqref{eq:KDSs} such that they are sorted by exterior and interior bonds and \emph{not} by ingoing and outgoing bonds, thus bringing \(\dirac_i\) into the form
\begin{multline} \label{eq:KDS:interiorexterior}
\dirac_i\ofx = \{ 
	( \begin{pmatrix}
	\left( \f_j\right)\\
	\left(\f_k\right)
	\end{pmatrix},
	\begin{pmatrix}
	\left(\e_j\right)\\
	\left(\e_k\right)
	\end{pmatrix} ) \in \R^{\dimBond \cdot \numB(i)} \times \R^{\dimBond \cdot \numB(i)} \mid  \\
\left( \left( \F_j\ofx \right)   \,  \left(\F_k\ofx\right)  \right) 
\begin{pmatrix}
	\left(\varepsilon(j) \f_j\right)\\
	\left(\varepsilon(k) \f_k\right)
	\end{pmatrix} + \\
\left( \left( \E_j\ofx \right)  \,  \left(\E_k\ofx\right)   \right)
	\begin{pmatrix}
	\left(\e_j\right) \\
	\left(\e_k\right)
\end{pmatrix}
= \0
	\},
\end{multline}
for all $j\in\B_\bgE(i), k\in \B_\bgI(i)$ where $\varepsilon\colon\B(i) \to \{-1,1\}$, $b \mapsto \varepsilon(b)$ is a sign function which is $1$ if $b \in \Bin(i)$ and $-1$ if $b \in \Bout(i)$.
For each $i\in\V_\bgI$, we define $\f_i^\bgIC \define (\varepsilon(k) \f_k)$ and $\e_i^\bgIC \define (\e_k)$ for all $k\in \B_\bgI(i)$.\footnote{The ``IC'' refers to ``interconnection''.} 
Furthermore, we write $\f^\bgIC \define (\f_i^\bgIC)$ and $\e^\bgIC \define (\e_i^\bgIC)$ for all $i\in\V_\bgI$. 
%
Each interior bond is incident to \emph{two} interior elements. 
Thus, for each $k\in\B_\bgI$ the flow $\f_k$ appears exactly twice in $\f^\bgIC$: once with a positive sign and once with a negative sign. 
Analogously, for each $k\in\B_\bgI$ the effort $\e_k$ appears exactly twice in $\e^\bgIC$, both times with a positive sign. 
Let us equate these variables appearing twice by setting
\begin{multline} \label{eq:equating:flowsefforts}
\begin{pmatrix}
	\I_{N \numB_\bgI} & \I_{N \numB_\bgI} \\
	\0_{N \numB_\bgI} & \0_{N \numB_\bgI}
	\end{pmatrix}
\begin{pmatrix*}[r]
	(\f_k)\\
	-(\f_k)
	\end{pmatrix*}
	+\\+
\begin{pmatrix*}[r]
	\0_{N \numB_\bgI} & \0_{N \numB_\bgI} \\
	\I_{N \numB_\bgI} & -\I_{N \numB_\bgI} 
	\end{pmatrix*}
\begin{pmatrix}
	(\e_k)\\
	(\e_k)
\end{pmatrix}
= \0
\end{multline}
for all $k \in \B_\bgI$. By permutations, we rearrange the entries of the vectors in \eqref{eq:equating:flowsefforts} such that they are in the same order as in $\f^\bgIC$ and $\e^\bgIC$. Furthermore, we rename the columns of the resulting matrices according to their affiliation to elements of $\V_\bgI$. 
The equation system is then of the form 
\begin{equation} \label{eq:IC:finalform}
\left( \F_i^\bgIC \right)
\left( \f_i^\bgIC \right)
+ 
\left( \E_i^\bgIC \right)
\left( \e_i^\bgIC \right)
= \0, \quad
\forall i\in\V_\bgI
\end{equation}
with the matrices $\F_i^\bgIC, \E_i^\bgIC \in \R^{2\dimBond \numB_\bgI \times \dimBond \numB_\bgI(i)}$.\footnote{By $2 \numB_\bgI = \sum_{i\in\V_\bgI} \numB_\bgI(i)$, the sizes of the matrices in \eqref{eq:equating:flowsefforts} and \eqref{eq:IC:finalform} are equal.} 
Let us define the vector space
\begin{equation} \label{eq:IC:DS}
\dirac_\bgIC =  
\{ (\f^\bgIC,\e^\bgIC) \in \R^{2 \dimBond \numB_\bgI} \times \R^{2 \dimBond  \numB_\bgI} \mid \eqref{eq:IC:finalform}~\text{holds} \}. 
\end{equation}
\begin{prop}
	$\dirac_\bgIC$ in~\eqref{eq:IC:DS} is a Dirac structure.
\end{prop}
\begin{pf}
The matrices in \eqref{eq:equating:flowsefforts} satisfy \eqref{eq:dirac:kernel:cond} and can thus be related to a Dirac structure. By a permutation matrix $\mat{T} \in\set{O}(2 \dimBond \numB_\bgI)$ we can reorder the entries of the vectors of \eqref{eq:equating:flowsefforts} to obtain \eqref{eq:IC:finalform}. 
By Corollary~\ref{cor:equivalentDS}, this proves \eqref{eq:IC:DS} to be a constant Dirac structure. 
\qed \end{pf}
Following the terminology of \cite{batlle2011}, \eqref{eq:IC:DS} is an \emph{interconnection Dirac structure} of the Dirac structures~\eqref{eq:KDS:interiorexterior}.
We now have all the required tools to compose the Dirac structures from \eqref{eq:KDSs} into one single Dirac structure. 
%
%

\begin{lem}[\cite{batlle2011}] \label{lemma:composition}
Consider $\n_\bgI$ Dirac structures of the form \eqref{eq:KDS:interiorexterior}. Furthermore, consider a corresponding interconnection Dirac structure of the form \eqref{eq:IC:DS}. Define a full-rank matrix $\mat{\Gamma}^\T \ofx \in \R^{2\dimBond \numB_\bgI \times \dimBond(2m_\bgI + \numB_\bgE)}$ as a $(1\times \numV_\bgI)$ block matrix $\mat{\Gamma}^\T\ofx = (\mat{\Gamma}_i^\T\ofx)$ of matrices $\mat{\Gamma}_i^\T\ofx \in \R^{2\dimBond \numB_\bgI \times \dimBond \numB(i)}$ for all $i\in\V_\bgI$ with
\begin{equation} \label{eq:KDS:Mi}
\mat{\Gamma}_i^\T\ofx = 
\F_i^\bgIC \left( \E_k\ofx \right)^\T +
\E_i^\bgIC \left( \F_k\ofx \right)^\T,~ \forall k\in\B_\bgI(i).
\end{equation}	
Choose a matrix $\mat{\Lambda} \ofx \in \R^{\dimBond \numB_\bgE \times \dimBond(2m_\bgI + \numB_\bgE)}$ such that $\img(\mat{\Lambda}^\T\ofx) = \ker(\mat{\Gamma}^\T\ofx)$ for all $\x\in\X$. Since $\rank(\mat{\Gamma}^\T\ofx) = 2 \dimBond \numB_\bgI$ for all $\x\in\X$, we have $\dim (\ker (\mat{\Gamma}^\T\ofx)) = \dimBond \numB_\bgE$ and such a matrix $\mat{\Lambda}\ofx$ always exists. 
Matrix $\mat{\Lambda}\ofx$ can be written as a $(1\times n_\bgI)$ block matrix $(\mat{\Lambda}_i\ofx)$ of matrices $\mat{\Lambda}_i\ofx \in \R^{\dimBond \numB_\bgE \times \dimBond \numB(i)}$ for all $i\in\V_\bgI$. Then the composite Dirac structure relates the flows $\f_j$ and efforts $\e_j$ of only the exterior bonds $j\in\B_\bgE$ and is of the form \eqref{eq:ds:kernel:blocks}: 
\begin{multline} \label{eq:singleKDS}
\dirac\ofx =  \{ 
	\left( (\f_j),(\e_j) \right) \in \R^{\dimBond \numB_\bgE} \times \R^{\dimBond \numB_\bgE} \mid \\
 \underbrace{\left(\mat{\Lambda}_i\ofx \left(\F_j\ofx\right) \right)}_{\definer \F \, \ofx} \left(\f_j\right) + 
\underbrace{ \left(\mat{\Lambda}_i\ofx \left(\E_j\ofx\right) \right)}_{\definer \E \, \ofx} \left(\e_j\right) 
= \0
	\},
\end{multline} 
for all $j\in\B_\bgE(i)$, $i\in\V_\bgI$. 
\end{lem}
\begin{pf}
	The proof for the more general case of any interconnection Dirac structure can be found in \cite{batlle2011}.
\end{pf}
\subsection{Explicit representation of the Dirac structure} \label{subsec:KDS2IODS}
In the previous section, we showed that it is always possible to determine a single \emph{implicit} Dirac structure~\eqref{eq:ds:kernel:blocks} describing the equations of the junction structure. 
In this section, we propose a constructive procedure for transferring the Dirac structure from an (implicit) kernel representation into an (explicit) input-output representation. 
As with the kernel representation, the input-output representation of a Dirac structure is not unique.  
In particular, not all explicit Dirac structures allow for a subsequent derivation of an explicit \phs~with Property~\ref{req:inputs_outputs}.
The inputs and outputs of an explicit \phs~are determined by the inputs and outputs of the underlying explicit Dirac structure. 
Thus, based on Property~\ref{req:inputs_outputs} we deduce the following property. 
\begin{req}\label{req:iods:inputs_outputs}
Let $\B_\alpha = \cup_{i\in\V_\alpha} \B(i)$ for $\alpha\in\{\bgSf,\bgSe\}$. The input vector of the explicit Dirac structure has to include $(\f_j)$, $(\e_k)$ while the output vector has to include $(\e_j)$, $(\f_k)$ for all $j\in\B_\bgSf$, $k\in\B_\bgSe$. 
\end{req}
In the sequel, we aim at an explicit representation of \eqref{eq:ds:kernel:blocks} that has Property~\ref{req:iods:inputs_outputs}. 
Necessary and sufficient conditions for the existence of such an explicit representation will be provided. \\
%
Given a Dirac structure in kernel representation~\eqref{eq:ds:kernel:blocks}. 
For the sake of notation, let us introduce
\sca{possible space save: inline formula}
\begin{subequations}
\begin{align}
	\F_{\bgC\bgR}\ofx &\define \left(\F_{\bgC}\ofx~\F_{\bgR}\ofx \right), \label{eq:FCR} \\ 
	\E_{\bgC\bgR}\ofx &\define \left(\E_{\bgC}\ofx~\E_{\bgR}\ofx \right), \label{eq:ECR}
\end{align}
\end{subequations}
as well as $\f_{\bgC\bgR} \define \left(\f_{\bgC}^\T~\f_{\bgR}^\T \right)^\T$ and $\e_{\bgC\bgR} \define \left(\e_{\bgC}^\T~\e_{\bgR}^\T \right)^\T.$ 
\begin{assum} \label{ass:independentsources}
The matrices in \eqref{eq:ds:kernel:blocks} fulfill \eqref{eq:theo:EF:fullrank}. 
\end{assum}
%
%
Based on Assumption~\ref{ass:independentsources} we can now state the following lemma. 
\begin{lem}\label{lemma:kds2iods}
Consider the Dirac structure~\eqref{eq:ds:kernel:blocks}. 
Let Assumption~\ref{ass:independentsources} hold. The Dirac structure can be formulated in an input-output representation with Property~\ref{req:iods:inputs_outputs}:  
\begin{multline}
\label{eq:dirac:io}
\dirac = \{ 
(
\begin{pmatrix}
	\f_{\bgC\bgR} \\
	\f_\bgSf \\
	\f_\bgSe \\
	\end{pmatrix},
	\begin{pmatrix}
	\e_{\bgC\bgR} \\
	\e_\bgSf \\
	\e_\bgSe \\
	\end{pmatrix}
)
	\in \R^{\dimBond\numV_\bgE} \times \R^{\dimBond\numV_\bgE}
	\mid \\
	\begin{pmatrix}
	\vecy_{\bgC\bgR} \\
	\vecy_\bgP \\
	\end{pmatrix} = 
	\matZ\ofx
	\begin{pmatrix}
	\vecu_{\bgC\bgR} \\
	\vecu_\bgP \\
	\end{pmatrix}
	\},
\end{multline}
where $\matZ\ofx$ is skew-symmetric for all $\x \in \X$ and
\begin{subequations}
\label{eq:dirac:io:vector:splitting}
\begin{alignat}{3}
	\vecu_{\bgC\bgR} & = \begin{pmatrix*}[r] \e_{\bgC\bgR,1} \\  -\! \f_{\bgC\bgR,2}	\end{pmatrix*}, 	\label{eq:dirac:io:vector:splitting:u}
	&\qquad&
	\vecu_\bgP & = \begin{pmatrix} \f_\bgSf	\\ \e_\bgSe \end{pmatrix}, 
	\\
	\vecy_{\bgC\bgR} & = \begin{pmatrix*}[r] - \f_{\bgC\bgR,1} \\ \e_{\bgC\bgR,2}	\end{pmatrix*},	\label{eq:dirac:io:vector:splitting:y}
	 &&
	\vecy_\bgP & = \begin{pmatrix} \e_\bgSf \\ \f_\bgSe \end{pmatrix}.
\end{alignat}
\end{subequations} 
The matrix $\matZ\ofx$ exists for all $\x \in \X$ and is given by:
\begin{equation} \label{eq:formula:Z}
	\begin{split}
		\mat{Z}\ofx &=
		\left( \F_{\bgC\bgR,1}\ofx ~ \E_{\bgC\bgR,2}\ofx ~ \E_\bgSf\ofx ~ \F_\bgSe\ofx \right)^{-1}
		\\
		&\phantom{={}}\cdot  \left( \E_{\bgC\bgR,1}\ofx ~ \F_{\bgC\bgR,2}\ofx ~ \F_\bgSf\ofx ~ \E_\bgSe\ofx \right)
		.
	\end{split}
\end{equation}
The matrices in \eqref{eq:formula:Z} can be obtained from splitting (possibly after some permutations) $\F_{\bgC\bgR} \ofx$ by \eqref{eq:FCR} into $\left( \F_{\bgC\bgR,1}\ofx\, \F_{\bgC\bgR,2}\ofx \right)$ such that
\begin{alignat}{2}\label{eq:kds2iods:splitting}
	&(i) && \left(\F_{\bgC\bgR,1}\ofx \, \E_\bgSf\ofx \, \F_\bgSe\ofx \right)~\text{has full column rank} \notag \\
	&(ii) && ~\rank \left(\F_{\bgC\bgR,1}\ofx ~ \E_\bgSf\ofx ~ \F_\bgSe\ofx \right) =  \notag \\
	& && \qquad \rank \left( \F_{\bgC\bgR}\ofx ~ \E_\bgSf\ofx ~ \F_\bgSe\ofx \right)
\end{alignat}
for all $\x\in\X$. According to the manner in which $\F_{\bgC\bgR} \ofx$ is split, we partition $\E_{\bgC\bgR} \ofx$ from \eqref{eq:ECR} into $\left(\E_{\bgC\bgR,1}\ofx \, \E_{\bgC\bgR,2}\ofx \right)$. 
In the same way, we split $\f_{\bgC\bgR}$ and $\e_{\bgC\bgR}$. 
\end{lem}
\begin{rem}
Note that the above lemma is true for any decomposition of \(\F_{\bgC\bgR}\) such that \eqref{eq:kds2iods:splitting} is fulfilled.
However, we choose \(\F_{\bgC\bgR,1}\) such that the number of columns originating from \(\F_{\bgC}\) is as large as possible since this is more useful for the subsequent derivation of an explicit \phs.
\end{rem}
\begin{pf}
%
%
Let Assumption~\ref{ass:independentsources} hold. For the sake of readability, we omit the argument $\x$ and the supplement ``for all $\x\in\X$'' in this proof. 
We apply the ideas from \cite[Theorem~4]{bloch1999} to show that we can always find decompositions $\left( \F_{\bgC\bgR,1}, \F_{\bgC\bgR,2} \right)$ and $\left( \E_{\bgC\bgR,1}, \E_{\bgC\bgR,2} \right)$ of $\F_{\bgC\bgR}$ and $\E_{\bgC\bgR}$ such that $\rank \left( \F_{\bgC\bgR,1} ~ \E_{\bgC\bgR,2} ~ \E_\bgSf ~ \F_\bgSe \right) = \dimBond\numV_\bgE$ holds. 
%
%
%
Choose a decomposition of \(\F_{\bgC\bgR}\) (possibly after some permutations) such that the conditions in~\eqref{eq:kds2iods:splitting} are fulfilled.
%
Next, split \(\E_{\bgC\bgR}\) according to the decomposition chosen for \(\F_{\bgC\bgR}\) into \(\E_{\bgC\bgR} = \left( \E_{\bgC\bgR,1} ~ \E_{\bgC\bgR,2} \right) \).
By \eqref{eq:kds2iods:splitting}, the matrix \((\F_{\bgC\bgR,1}~\E_\bgSf~\F_\bgSe)\) has full column rank. Thus, its adjoint \((\F_{\bgC\bgR,1}~\E_\bgSf~\F_\bgSe)^\T\) is surjective.
In particular we have
\begin{multline}
	\label{eq:image E+F 1}
	\img \begin{pmatrix} \E_{\bgC\bgR,1} & \F_\bgSf & \E_\bgSe \end{pmatrix}
	\\
	= \img \left(\begin{pmatrix} \E_{\bgC\bgR,1} & \F_\bgSf & \E_\bgSe \end{pmatrix}
	\cdot
	\begin{pmatrix} \F_{\bgC\bgR,1} & \E_\bgSf & \F_\bgSe\end{pmatrix}^\T
	\right)
	\\
	= \img \left(\E_{\bgC\bgR,1}\F_{\bgC\bgR,1}^\T + \F_\bgSf \E_\bgSf^\T + \E_\bgSe\F_\bgSe^\T \right)
	.
\end{multline}
Equation~\eqref{eq:diracKern} is
\begin{equation}
	\0
	= \E\F^\T + \F\E^\T 
	= \smashoperator{\sum_{\alpha\in\{(\bgC\bgR,1),(\bgC\bgR,2),\bgSf,\bgSe \}}} \left(\E_\alpha \F_\alpha^\T + \F_\alpha\E_\alpha^\T\right)
\end{equation}
from which follows
\begin{equation}
	\label{eq:image E+F 2}
	\begin{split}
	\MoveEqLeft
	\img \left(\E_{\bgC\bgR,1}\F_{\bgC\bgR,1}^\T + \F_\bgSf \E_\bgSf^\T + \E_\bgSe\F_\bgSe^\T \right)
	\\&= \begin{multlined}[t]
		\img\left(
			\F_{\bgC\bgR,1}\E_{\bgC\bgR,1}^\T +
			\E_{\bgC\bgR,2}\F_{\bgC\bgR,2}^\T +
		\right.
		\\+
		\left.
			\F_{\bgC\bgR,2}\E_{\bgC\bgR,2}^\T +
			\E_\bgSf\F_\bgSf^\T +
			\F_\bgSe\E_\bgSe^\T
		\right)
	\end{multlined}
	\\&\subseteq \begin{multlined}[t]
		\img\Bigl(
		\F_{\bgC\bgR,1}\E_{\bgC\bgR,1}^\T ~~
		\E_{\bgC\bgR,2}\F_{\bgC\bgR,2}^\T ~~
		\\
		\F_{\bgC\bgR,2}\E_{\bgC\bgR,2}^\T ~~
		\E_\bgSf\F_\bgSf^\T ~~
		\F_\bgSe\E_\bgSe^\T
		\Bigr)
	\end{multlined}
	\\&\subseteq \img\begin{pmatrix}
		\F_{\bgC\bgR,1} &
		\E_{\bgC\bgR,2} &
		\F_{\bgC\bgR,2} &
		\E_\bgSf &
		\F_\bgSe
	\end{pmatrix}
	\\&\stackrel{\mathclap{\eqref{eq:kds2iods:splitting}}}{=} \img\begin{pmatrix}
		\F_{\bgC\bgR,1} &
		\E_{\bgC\bgR,2} &
		\E_\bgSf &
		\F_\bgSe
	\end{pmatrix}
	.
	\end{split}
\end{equation}
Combining \eqref{eq:image E+F 1} and~\eqref{eq:image E+F 2} we can derive
\begin{equation}
	\begin{split}
		\MoveEqLeft\img \begin{pmatrix} \F & \E \end{pmatrix}
	\\&=
	\begin{multlined}[t]
	\img \begin{pmatrix} \E_{\bgC\bgR,1} & \F_\bgSf & \E_\bgSe \end{pmatrix}
	+\\+
	\img\begin{pmatrix}
		\F_{\bgC\bgR,1} &
		\E_{\bgC\bgR,2} &
		\F_{\bgC\bgR,2} &
		\E_\bgSf &
		\F_\bgSe
	\end{pmatrix}
	\end{multlined}
	\\&\subseteq
	\img\begin{pmatrix}
		\F_{\bgC\bgR,1} &
		\E_{\bgC\bgR,2} &
		\E_\bgSf &
		\F_\bgSe
	\end{pmatrix}
	\\&\subseteq
	\img \begin{pmatrix} \F & \E \end{pmatrix}.
	\end{split}
\end{equation}
Thus, equality holds in the above formula and we have
\begin{multline}
	\label{eq:rank:YYY}
	\rank \begin{pmatrix} \F_{\bgC\bgR,1} & \E_{\bgC\bgR,2} & \E_\bgSf & \F_\bgSe \end{pmatrix}
	\\
	= \rank \begin{pmatrix} \F & \E \end{pmatrix}
	\stackrel{\eqref{eq:diracRank}}{=} \dimBond\numV_\bgE.
\end{multline}
Hence, the square matrix \((\F_{\bgC\bgR,1} ~ \E_{\bgC\bgR,2} ~ \E_\bgSf ~ \F_\bgSe)\) has full rank and is invertible. 
As shown in \cite{bloch1999} and \cite{golo2000}, under the above rank condition~\eqref{eq:rank:YYY} the kernel representation~\eqref{eq:ds:kernel:blocks} can be formulated as the input-output representation~\eqref{eq:dirac:io} with
\begin{equation} 
	\mat{Z} = -\mat{Z}^\T = \begin{multlined}[t]
		- \begin{pmatrix} \F_{\bgC\bgR,1} & \E_{\bgC\bgR,2} & \E_\bgSf & \F_\bgSe \end{pmatrix}^{-1}
		\cdot\\\cdot
		\begin{pmatrix} \E_{\bgC\bgR,1} & \F_{\bgC\bgR,2} & \F_\bgSf & \E_\bgSe \end{pmatrix}
	.
	\end{multlined}
\end{equation}
\qed
\end{pf}
As can be seen in \eqref{eq:dirac:io:vector:splitting}, the flows and efforts corresponding to elements of type $\bgSf$ and $\bgSe$ are assigned as inputs and outputs of the explicit Dirac structure in a \emph{fixed} manner. By this fixed assignment, \eqref{eq:dirac:io} has Property~\ref{req:iods:inputs_outputs}. In contrast, the flows and efforts corresponding to elements of type $\bgC$ and $\bgR$ may be freely designated as inputs or outputs as long as \eqref{eq:kds2iods:splitting} is fulfilled. 
In the next two propositions, we analyse the result of Lemma~\ref{lemma:kds2iods} more in detail.  
\begin{prop}\label{prop:uniqueness:Z}
For any given order of the variables in \eqref{eq:dirac:io:vector:splitting}, the matrix \(\matZ\ofx\) in \eqref{eq:dirac:io} is unique. This statement is independent of Assumption~\ref{ass:independentsources}.
\end{prop}
%
\begin{pf}
%
\newcommand{\ff}{\begin{pmatrix}
	-\f_{\bgC,1} \\
	-\f_{\bgC,2} \\
	\f_\bgSf \\
	\f_\bgSe \\
	-\f_{\bgR,1} \\
	-\f_{\bgR,2} \\
	\end{pmatrix}}
\newcommand{\ee}{\begin{pmatrix}
	\e_{\bgC,1} \\
	\e_{\bgC,2} \\
	\e_\bgSf \\
	\e_\bgSe \\
	\e_{\bgR,1} \\
	\e_{\bgR,2} \\
	\end{pmatrix}}
\newcommand{\yy}{\begin{pmatrix}
	\vecy_\bgC \\
	\vecy_\bgP \\
	\vecy_\bgR
	\end{pmatrix}}
\newcommand{\uu}{\begin{pmatrix}
	\vecu_\bgC \\
	\vecu_\bgP \\
	\vecu_\bgR
	\end{pmatrix}}
The idea is to show that \(\dirac\) is linearly isomorphic to \(\R^{\dimBond\numV_\bgE}\) (i.e.\ isomorphic as vector spaces) and thus $\matZ$ is unique. For the sake of releasing notational burden, we will suppress the argument \(\x\) to the matrices during the proof and use the following notation: 
%
\begin{subequations} \label{eq:shorthand:uyef}
\begin{align}
	\vecu &= \left(\vecu_{\bgC\bgR}^\T~\vecu_\bgP^\T\right)^\T, \label{eq:shorthand:uyef:u}\\
	\vecy & = \left(\vecy_{\bgC\bgR}^\T~\vecy_\bgP^\T\right)^\T, \label{eq:shorthand:uyef:y} \\
	\f & = \left(\f_{\bgC\bgR,1}^\T ~	\f_{\bgC\bgR,2}^\T ~	\f_\bgSf^\T ~	\f_\bgSe^\T \right)^\T, \\
	\e &=	\left(\e_{\bgC\bgR,1}^\T ~	\e_{\bgC\bgR,2}^\T ~	\e_\bgSf^\T ~	\e_\bgSe^\T \right)^\T. 
	\end{align}
\end{subequations}
Let \(\matZ\) and \(\mat Z' \in \R^{\dimBond\numV_\bgE \times \dimBond\numV_\bgE}\) be two matrices fulfilling 
\begin{equation} \label{eq:lamma:Z=Z'}
	\vecy = \mat Z \vecu
\quad\text{and}\quad
	\vecy = \mat Z' \vecu. 
\end{equation}
Recall \eqref{eq:dirac:io:vector:splitting} and that \(\dim \dirac = \dimBond\numV_\bgE\).
As \(\vecy\) depends linearly on \(\vecu\), we have that \(\dirac\) is isomorphic to \(\R^{\dimBond\numV_\bgE}\) via $\R^{\dimBond\numV_\bgE} \to \dirac$, $\vecu \mapsto (\f,\e)$, 
\begingroup
\newcommand{\fff}{\begin{pmatrix}
	\f_{\bgC,1} \\
	\f_{\bgC,2} \\
	\f_\bgSf \\
	\f_\bgSe \\
	\f_{\bgR,1} \\
	\f_{\bgR,2} \\
	\end{pmatrix}}
\newcommand{\eee}{\begin{pmatrix}
	\e_{\bgC,1} \\
	\e_{\bgC,2} \\
	\e_\bgSf \\
	\e_\bgSe \\
	\e_{\bgR,1} \\
	\e_{\bgR,2} \\
	\end{pmatrix}}
where $\vecy = \mat Z\vecu$
and via $\dirac \to \R^{\dimBond\numV_\bgE}$, $(\e,\f) \mapsto \vecu$.
\endgroup
From \eqref{eq:lamma:Z=Z'} it follows that $\mat Z \vecu = \mat Z' \vecu$
and thus \(\mat Z = \mat Z'\) as \(\vecu\) ranges over all of \(\R^{\dimBond\numV_\bgE}\).
\qed
\end{pf}
Note that the uniqueness of \(\matZ\ofx\) in Proposition~\ref{prop:uniqueness:Z} is restricted to the case of a certain arrangement of variables. In particular, Proposition~\ref{prop:uniqueness:Z} does \emph{not} imply the uniqueness of an input-output representation in general.    
\begin{prop} \label{prop:wellposedness:necc+suff}
Assumption~\ref{ass:independentsources} is a \emph{necessary and sufficient condition} for the existence of an input-output representation of \eqref{eq:ds:kernel:blocks} which has Property~\ref{req:iods:inputs_outputs}. This statement is true independent of the specific realisation of $\F\ofx$ and $\E\ofx$ in \eqref{eq:ds:kernel:blocks} (cf. Remark~\ref{rem:uniqueness:FE}).
\end{prop}
%
\begin{pf}
\newcommand{\ff}{\begin{pmatrix}
	-\f_{\bgC,1} \\
	-\f_{\bgC,2} \\
	\f_\bgSf \\
	\f_\bgSe \\
	-\f_{\bgR,1} \\
	-\f_{\bgR,2} \\
	\end{pmatrix}}
\newcommand{\ee}{\begin{pmatrix}
	\e_{\bgC,1} \\
	\e_{\bgC,2} \\
	\e_\bgSf \\
	\e_\bgSe \\
	\e_{\bgR,1} \\
	\e_{\bgR,2} \\
	\end{pmatrix}}
\newcommand{\yy}{\begin{pmatrix}
	\vecy_\bgC \\
	\vecy_\bgP \\
	\vecy_\bgR
	\end{pmatrix}}
\newcommand{\uu}{\begin{pmatrix}
	\vecu_\bgC \\
	\vecu_\bgP \\
	\vecu_\bgR
	\end{pmatrix}}
From the proof of Lemma~\ref{lemma:kds2iods} it follows that Assumption~\ref{ass:independentsources} is a \emph{sufficient condition} for transferring \eqref{eq:ds:kernel:blocks} into an input-output representation with property~\ref{req:iods:inputs_outputs}. 
So it is left to show that the assumption is \emph{necessary}.
To this end, we use the uniqueness of $\matZ\ofx$ from Proposition~\ref{prop:uniqueness:Z}.
For the sake of brevity, we neglect the argument $\x$ and the supplement ``for all $\x\in\X$'' in this proof.
Moreover, we use the notation from \eqref{eq:shorthand:uyef:u}, \eqref{eq:shorthand:uyef:y} and we give a shorthand to two matrices:
\sca{possible space save: inline formula}
\begin{subequations} \label{eq:lemma:necessary:0}
\begin{align}
	\mat X &= \begin{pmatrix} \F_{\bgC\bgR,1} & \E_{\bgC\bgR,2} & \E_\bgSf & \F_\bgSe \end{pmatrix} \in \R^{\dimBond\numV_\bgE \times \dimBond\numV_\bgE}, \label{eq:lemma:necessary:0a}\\
	\mat Y &= \begin{pmatrix} \E_{\bgC\bgR,1} & \F_{\bgC\bgR,2} & \F_\bgSf & \E_\bgSe \end{pmatrix} \in \R^{\dimBond\numV_\bgE \times \dimBond\numV_\bgE}.
\end{align}
\end{subequations}
Assume we can write \(\dirac\) in both forms \eqref{eq:ds:kernel:blocks} and \eqref{eq:dirac:io}. 
Moreover, Assumption~\ref{ass:independentsources} is fulfilled if \(\mat X\) has full rank. 
Note that in the situation of Lemma~\ref{lemma:kds2iods} we have \(\matZ = -\mat X^{-1}\mat Y\) which gives us a hint that we should prove and use \(\mat X \matZ = -\mat Y\) along the way.
As an element \((\f,\e)\) of \(\dirac\) fulfills the equations in \eqref{eq:ds:kernel:blocks}, we have
%
\begin{multline}
	\mat F  
	\begin{pmatrix}
	-\f_{\bgC\bgR,1}^\T &
	-\f_{\bgC\bgR,2}^\T &
	\f_\bgSf^\T &
	\f_\bgSe^\T &
	\end{pmatrix}^\T \\ + 
	\mat E 
	\begin{pmatrix}
	\e_{\bgC\bgR,1}^\T &
	\e_{\bgC\bgR,2}^\T &
	\e_\bgSf^\T &
	\e_\bgSe^\T &
	\end{pmatrix}^\T
	= \0,
\end{multline}
or equivalently after reordering 
\begin{equation}
	\label{eq:lemma:necessary:1}
	\mat X \vecy = - \mat Y \vecu.
\end{equation}
The same element \((\f,\e)\) also fulfills \eqref{eq:dirac:io}, i.e. we have $\vecy = \matZ \vecu$, 
where $\matZ$ is unique according to Proposition~\ref{prop:uniqueness:Z}. By multiplying from the left with $\mat X$ we obtain
\begin{equation}
	\label{eq:lemma:necessary:2}
	\mat X \vecy = \mat X \matZ \vecu
	.
\end{equation}
Combining \eqref{eq:lemma:necessary:1} and \eqref{eq:lemma:necessary:2} yields
\sca{possible space save: inline formula}
\begin{equation}	\label{eq:lemma:necessary:3}
	\mat X\matZ \vecu = - \mat Y \vecu,
\end{equation}
establishing \(\mat X \matZ = - \mat Y\), since \(\vecu\) ranges over all of \(\R^{\dimBond\numV_\bgE}\). \\
Let us now investigate the rank of $\mat X$. First, note that \(\img \mat X = \img(\mat X~\mat X \mat Z)\) as $\img\mat{X}\mat{Z}\subseteq\img\mat{X}$. 
From this the statement that \(\mat X\) has full rank follows:
\sca{possible space save: inline formula}
\begin{multline} \label{eq:fullrank:necc}
	\rank \mat X
	= \rank \begin{pmatrix}\mat X ~ \mat X\mat Z\end{pmatrix}
	\stackrel{\eqref{eq:lemma:necessary:3}}{=} \rank \begin{pmatrix}\mat X ~ -\mat Y\end{pmatrix}
	\\
	\stackrel{\eqref{eq:lemma:necessary:0}}{=} \rank \begin{pmatrix} \mat F ~ \mat E\end{pmatrix}
	\stackrel{\eqref{eq:dirac:def:ii}}{=} \dimBond\numV_\bgE.
\end{multline}
Note that \eqref{eq:fullrank:necc} holds for any realisation of $\mat F$ and $\mat E$. 
Moreover, every submatrix in \eqref{eq:lemma:necessary:0a} must have full column rank.
In particular Assumption~\ref{ass:independentsources} holds.
\qed
\end{pf}

So far, we presented a method which allows to convert the Dirac structure~\eqref{eq:ds:kernel:blocks} to an explicit form~\eqref{eq:dirac:io}. 
In the sequel, we consider an important special case of \eqref{eq:dirac:io} which will pave the way to a port-Hamiltonian formulation of the bond graph. The special case is characterised by the following assumption. 
%
%
\begin{assum} \label{ass:storagesandsources:independent}
For all $\x\in\X$ the matrices in \eqref{eq:ds:kernel:blocks} fulfill \eqref{eq:FEF:fullrank}.
\end{assum}
Note that Assumption~\ref{ass:independentsources} is necessary for Assumption~\ref{ass:storagesandsources:independent}.
In the subsequent corollary, we make use of Assumption~\ref{ass:storagesandsources:independent} and address an important special case of Lemma~\ref{lemma:kds2iods}. 
\begin{cor} \label{cor:kds2iods:specialform}
Given the Dirac structure~\eqref{eq:ds:kernel:blocks}. 
Let Assumption~\ref{ass:storagesandsources:independent} hold. 
The Dirac structure~\eqref{eq:ds:kernel:blocks} can then be formulated in the input-output representation~\eqref{eq:dirac:io:specialform}. 
Moreover, Assumption~\ref{ass:storagesandsources:independent} is necessary and sufficient for the existence of \eqref{eq:dirac:io:specialform} with vectors as in \eqref{eq:dirac:io:u} and \eqref{eq:dirac:io:y}.
\end{cor}
\begin{pf}
The proof of Corollary~\ref{cor:kds2iods:specialform} follows directly from Lemma~\ref{lemma:kds2iods} under Assumption~\ref{ass:storagesandsources:independent}, which also shows that Assumption~\ref{ass:storagesandsources:independent} is a sufficient condition.
The proof for the necessity of the assumption is the same as the one given for Proposition~\ref{prop:wellposedness:necc+suff}.
\qed
\end{pf}
%
%
Lemma~\ref{lemma:kds2iods} provides a practical procedure for transferring the Dirac structure from a kernel representation \eqref{eq:ds:kernel:blocks} into an input-output representation \eqref{eq:dirac:io} with Property~\ref{req:iods:inputs_outputs}. Assumption~\ref{ass:independentsources} is proven to be necessary and sufficient for the existence of such a representation. In Corollary~\ref{cor:kds2iods:specialform}, we considered an important special case of Lemma~\ref{lemma:kds2iods}, which will be used to derive an explicit \phs~from the bond graph in the next section. 
%
\subsection{Formulation of an explicit port-Hamiltonian system} \label{subsec:IODS2PHS}
In the previous section, we showed that under certain conditions an explicit representation of the Dirac structure~\eqref{eq:ds:kernel:blocks} can be obtained. 
Hereby, the inputs and outputs of the explicit representation are chosen under consideration of Property~\ref{req:inputs_outputs}.  
In this section, we merge the explicit representation of the Dirac structure with the constitutive relations of storages and resistors to obtain an explicit \phs~\eqref{eq:phs} that has Property~\ref{req:inputs_outputs}. For this, let us make the following assumption. 
\begin{assum} \label{ass:resistiveRelation}
The resistive relation \eqref{eq:resistivePort} can be reorganised as in \eqref{eq:ass:R-port:theo}.
\end{assum}
\noindent The negative sign in \eqref{eq:ass:R-port:theo} accounts for the opposite signs of the flows in the vectors $\left(\f_\bgR, \e_\bgR \right)$ and $\left(\vecu_\bgR,\vecy_\bgR \right)$ (see~\eqref{eq:dirac:io:u} and \eqref{eq:dirac:io:y}). Before we formulate the bond graph as \phs, we need one more prerequisite lemma, which ensures the existence of \(\tilde{\mat{K}}\) in \eqref{eq:Ktilde}.
%
\begin{lem} \label{lemma:regularityK}
Let $\mat{X}, \mat{Y} \in \R^{p \times p}$ with $\mat{X}=\mat{X}^\T \possemidef$ and $\mat{Y}=-\mat{Y}^\T$. Then, the matrix \( \mat{K} \define \left( \I+\mat{X}\mat{Y} \right) \) is regular. In particular \(\mat{K}^{-1}\) always exists.
\end{lem}
%
\begin{pf}
\label{proof:regularityK}
The idea of the proof is to show that 
(i) we can (without loss of generality) assume \(\mat{X}\) to be diagonal;
(ii) the matrix $\mat{K}$ is invertible as it has only non-zero eigenvalues.
For (ii) we investigate first the case of $\mat{X}$ being positive definite. Afterwards, we generalise to the case of $\mat{X}$ being positive semi-definite. \\
%
Indeed, since \(\mat{X}\) is a symmetric and real matrix, there exists (by the Spectral Theorem) an orthogonal matrix \(\mat{T} \in \set{O}(p)\) such that \(\mat T \mat{X}\mat T^\T\) is diagonal.
Moreover, \(\I+\mat{X}\mat{Y}\) is invertible if and only if \(\mat T(\I+\mat{X}\mat{Y})\mat T^\T = \I + (\mat{T}\mat{X}\mat{T}
^\T) (\mat{T}\mat{Y}\mat{T}^\T) = \I + \tilde{\mat{X}}\tilde{\mat{Y}}\) is invertible, where \(\tilde{\mat{X}} = \mat{T}\mat{X}\mat{T}^\T\) is diagonal and positive semi-definite and \(\tilde{\mat{Y}} = \mat{T}\mat{Y}\mat{T}^\T\) is skew-symmetric. Thus, we can assume $\mat{X}$ to be diagonal in the remainder of the proof. \\
The matrix \(\I+\mat{X}\mat{Y}\) is regular if and only if 0 is not an eigenvalue of it, that is if \(-1\) is not an eigenvalue of \(\mat{X}\mat{Y}\).
We will show that \(\mat{X}\mat{Y}\) has at most \(0\) as \emph{real-valued} eigenvalue.
Throughout this proof we use \(\Spec(\mat{\cdot})\) to denote the (real) spectrum of a matrix. 
\\
Case 1: \(\mat{X}\) is positive definite.
Let \(\sqrt{\mat{X}}\) be a diagonal matrix which is a square root of \(\mat{X}\), i.e.~\(\sqrt{\mat{X}}\sqrt{\mat{X}} = \mat{X}\).
Such a matrix exists and is invertible since \(\mat{X}\) is diagonal and positive definite.
Because the spectrum of a matrix is invariant under conjugation, we have
\sca{possible space save: inline formula}
\begin{multline}
\Spec\left(\mat{X}\mat{Y}\right)
=\Spec\left(\sqrt{\mat{X}}^{-1}\mat{X}\mat{Y} \sqrt{\mat{X}}\right)
\\
=\Spec\left(\sqrt{\mat{X}}\mat{Y} \sqrt{\mat{X}}\right)
=\Spec\left(\sqrt{\mat{X}}\mat{Y} \sqrt{\mat{X}}^\T\right)
\subseteq\{0\}
,
\end{multline}
where the last inclusion holds since \(\sqrt{\mat{X}}\mat{Y} \sqrt{\mat{X}}^{\T}\) is real and skew-symmetric.
Thus, \(-1\) is not an eigenvalue of \(\mat{X}\mat{Y}\) and \(\I+\mat{X}\mat{Y}\) is invertible. \\
Case 2: \(\mat{X}\) is positive semi-definite.
By the same conjugation argument as at the beginning of the proof (this time with a permutation matrix) we may assume without loss of generality that \(\mat{X}\) is of the form
\sca{possible space save: inline formula}
\begin{equation}
	\mat{X} = \begin{pmatrix}\mat{X}' & \mat 0\\ \mat 0 & \mat 0\end{pmatrix},
\end{equation}
where \(\mat{X}' \in \R^{\ell\times\ell}\) is a positive definite diagonal matrix.
With the same block decomposition we write \(\mat{Y}\) as
\sca{possible space save: inline formula}
\begin{equation}
	\mat{Y} = \begin{pmatrix}\mat{Y}' & \mat{Y}'' \\ \ast & \ast \end{pmatrix},
	\qquad\text{where \(\mat{Y}' \in \R^{\ell\times\ell}\).}
\end{equation}
We have
\sca{possible space save: inline formula}
\begin{equation}
	\mat{X} \mat{Y} = \begin{pmatrix}\mat{X}'\mat{Y}' & \mat{X}'\mat{Y}'' \\ \mat 0 & \mat 0 \end{pmatrix}.
\end{equation}
Thus, \(\Spec(\mat{X} \mat{Y}) = \Spec(\mat{X}' \mat{Y}') \cup \Spec(\mat 0) \subseteq \{0\}\), where the last inclusion  uses Case 1 applied to \(\mat{X}' \mat{Y}'\).
Hence, \(\I+\mat{X}\mat{Y}\) is invertible.
\qed
\end{pf}
In the following, we use Lemma~\ref{lemma:regularityK} to merge the explicit Dirac structure~\eqref{eq:dirac:io:specialform} and the constitutive relations of storages~\eqref{eq:storagePort} and resistors~\eqref{eq:ass:R-port:theo} into an explicit \phs. 
%
\begin{lem} \label{lemma:iods2phs}
Given an explicit Dirac structure~\eqref{eq:dirac:io:specialform} and the constitutive relations of storage elements as in \eqref{eq:storagePort}. Suppose Assumption~\ref{ass:resistiveRelation} holds, which allows the constitutive relations of the resistive elements \eqref{eq:resistivePort} to be written as in \eqref{eq:ass:R-port:theo}. Equations~\eqref{eq:dirac:io:specialform}, \eqref{eq:storagePort}, and \eqref{eq:ass:R-port:theo} can be written as explicit input-state-output \phs~of the form \eqref{eq:phs} with state $\x$ and Hamiltonian $H\ofx$ from \eqref{eq:storagePort} and $\vecu = \vecu_\bgP$, $\vecy = \vecy_\bgP$. 

\end{lem}
\begin{pf}
The proof follows four steps: (i) we eliminate the resistive variables in \eqref{eq:dirac:io:specialform}; (ii) we decompose the structure obtained from (i) into symmetric and skew-symmetric parts; (iii) we substitute storage variables with \eqref{eq:storagePort}; (iv) we show that \eqref{eq:phs:definiteness} holds. Again, we omit the argument $\x$ and the supplement ``for all $\x\in\X$'' for all matrices in this proof.
\\
Substituting the second row from the linear equation system in \eqref{eq:dirac:io:specialform} into \eqref{eq:ass:R-port:theo} yields
\sca{possible space save: inline formula}
\begin{align}
\vecu_\bgR &= -\tilde{\matR} \matZ_{\bgC\bgR}^\T \vecu_\bgC +\tilde{\matR} \matZ_{\bgR\bgP}\vecu_\bgP - \tilde{\matR} \matZ_{\bgR\bgR} \vecu_\bgR \notag \\
\Leftrightarrow \vecu_\bgR &=  -\tilde{\mat{K}} \tilde{\matR} \matZ_{\bgC\bgR}^\T \vecu_\bgC + \tilde{\mat{K}} \tilde{\matR} \matZ_{\bgR\bgP} \vecu_\bgP
\label{eq:real:hio2phs:aux1} 
\end{align}
with $\tilde{\mat{K}}$ as in \eqref{eq:Ktilde}. Due to Lemma~\ref{lemma:regularityK}, $\tilde{\mat{K}}$ always exists. Inserting \eqref{eq:real:hio2phs:aux1}  into the first and third row from the linear equation system in  \eqref{eq:dirac:io:specialform} yields
\begin{multline}
\label{eq:real:hio2phs:aux3} 
	\begin{pmatrix}
	\vecy_\bgC \\
	\vecy_\bgP 
	\end{pmatrix}  =   \left[ 
	\begin{pmatrix*}[r]
	\matZ_{\bgC\bgC} 	& 
		-\matZ_{\bgC\bgP} 	\\
	 \matZ_{\bgC\bgP}^\T  &
	\matZ_{\bgP\bgP} 
	\end{pmatrix*} \right.  + \\
\left. \begin{pmatrix}
	\matZ_{\bgC\bgR} \\
	-\matZ_{\bgR\bgP}^\T
\end{pmatrix}
 \tilde{\mat{K}}\tilde{\matR}  
\begin{pmatrix}
	\matZ_{\bgC\bgR}^\T &
	-\matZ_{\bgR\bgP}
\end{pmatrix}
	\right]    
\begin{pmatrix} 
	\vecu_\bgC \\
	\vecu_\bgP 
	\end{pmatrix}.
\end{multline}
The first addend in the square bracket is a skew-symmetric matrix. The second addend can be decomposed into a skew-symmetric and a symmetric matrix by writing $2\tilde{\mat{K}}\tilde{\matR}$ as a sum of a skew-symmetric \(\mat{A}\) and a symmetric matrix \(\mat{B}\). Using this decomposition and $\tilde{\matR} = \tilde{\matR}^\T$, \eqref{eq:real:hio2phs:aux3} reads
\begin{multline}
\label{eq:real:hio2phs:aux4} 
	\begin{pmatrix}
	\vecy_\bgC \\
	\vecy_\bgP 
	\end{pmatrix} = 
	\begin{pmatrix*}[r]
	\matZ_{\bgC\bgC} 	& 
		-\matZ_{\bgC\bgP} 	\\
	 \matZ_{\bgC\bgP}^\T  &
	\matZ_{\bgP\bgP} 
	\end{pmatrix*} 
	\begin{pmatrix} 
	\vecu_\bgC \\
	\vecu_\bgP 
\end{pmatrix}
\\
	+ \frac{1}{2}
\begin{pmatrix}
	\matZ_{\bgC\bgR} \\
	-\matZ_{\bgR\bgP}^\T
\end{pmatrix}
\left( \mat{A} + \mat{B} \right)
\begin{pmatrix}
	\matZ_{\bgC\bgR}^\T &
	-\matZ_{\bgR\bgP}
\end{pmatrix} 
\begin{pmatrix} 
	\vecu_\bgC \\
	\vecu_\bgP 
\end{pmatrix},
\end{multline}
with $\mat{A}$ and $\mat{B}$ as in \eqref{eq:iods2phs:A} and \eqref{eq:iods2phs:B}, respectively. 
Equation \eqref{eq:real:hio2phs:aux4} can be written as 
\begin{equation}
\label{eq:real:hio2phs:aux5} 
	\begin{pmatrix}
	\vecy_\bgC \\
	\vecy_\bgP 
	\end{pmatrix} = \underbrace{
	\left[ 
	\begin{pmatrix}
	-\matJ & -\matG \\
	\matG^\T& \matM
	\end{pmatrix} \right.
	}_{=:\mat{Q}_{ss}}
	\tightplus \underbrace{\left.
\begin{pmatrix}
\matR	& \matP\\
\matP^\T & \matS
\end{pmatrix}
\right]}_{=:\mat{Q}_{s}}
\begin{pmatrix} 
	\vecu_\bgC \\
	\vecu_\bgP 
\end{pmatrix},
\end{equation}
with $\matJ$, $\matG$, $\matM$, $\matR$, $\matP$, $\matS$ as in \eqref{eq:real:phs_matrices_from_HIO_DS} and $\mat{Q}_{ss}=-\mat{Q}_{ss}^\T$, $\mat{Q}_{s}=\mat{Q}_{s}^\T$. Inserting \eqref{eq:storagePort} into \eqref{eq:real:hio2phs:aux5} then yields \eqref{eq:phs}.
Using the idea of \cite[p.~56]{van_der_schaft2014}, we eventually prove \eqref{eq:phs:definiteness}:
\begin{multline}
\begin{pmatrix} \vecu_\bgC^\T & \vecu_\bgP^\T \end{pmatrix} \mat{Q}_{s} \begin{pmatrix} \vecu_\bgC \\ \vecu_\bgP \end{pmatrix} = 
\begin{pmatrix} \vecu_\bgC^\T & \vecu_\bgP^\T \end{pmatrix} \left( \mat{Q}_{ss} + \mat{Q}_{s} \right) \begin{pmatrix} \vecu_\bgC \\ \vecu_\bgP \end{pmatrix}    \\ 
= \begin{pmatrix} \vecu_\bgC^\T & \vecu_\bgP^\T \end{pmatrix} \begin{pmatrix} \vecy_\bgC \\ \vecy_\bgP \end{pmatrix}
\stackrel{\eqref{eq:dirac:def:i}}{=} -\vecy_\bgR^\T \vecu_\bgR \stackrel{\eqref{eq:ass:R-port:theo}}{=} \vecy_\bgR^\T \tilde{\matR} \vecy_\bgR \geq 0.
\end{multline} \qed
\end{pf}
\begin{rem}\label{rem:literature:iods2phs}
The authors of \cite{donaire2009} derive an explicit \phs~\emph{without} feedthrough for the case $\matZ_{\bgP\bgP}\ofx = \0$, $\matZ_{\bgR\bgP}\ofx = \0$. In \cite{lopes2016}, the problem has been addressed for the special case $\matZ_{\bgR\bgR}\ofx  = \0$. Lemma~\ref{lemma:iods2phs} generalises the results of \cite{donaire2009} and \cite{lopes2016} to the case where all matrices of \eqref{eq:dirac:io:specialform} are potentially non-zero. \\
\end{rem}
%
\subsection{Necessary and sufficient conditions} \label{sec:proof:thm2}
Sections~\ref{subsec:JS2DSs} to~\ref{subsec:IODS2PHS} showed that if \eqref{eq:FEF:fullrank} and \eqref{eq:ass:R-port:theo} are fulfilled, an explicit port-Hamiltonian formulation of the bond graph can be obtained. 
Hence, under Property~\ref{req:inputs_outputs}, equations \eqref{eq:FEF:fullrank} and \eqref{eq:ass:R-port:theo} form together a sufficient condition for the existence of such an explicit \phs. 
Now it is left to show that \eqref{eq:theo:EF:fullrank} is a necessary condition for the existence of a port-Hamiltonian formulation that has Property~\ref{req:inputs_outputs}. \\
Property~\ref{req:inputs_outputs} implies Property~\ref{req:iods:inputs_outputs}. In Proposition~\ref{prop:wellposedness:necc+suff} we show that \eqref{eq:theo:EF:fullrank} is necessary (and sufficient) for formulating the junction structure equations as a Dirac structure in an explicit representation with Property~\ref{req:iods:inputs_outputs}.
In Lemma~\ref{lemma:iods2phs} it is shown, that the inputs and outputs of the explicit Dirac structure directly translate into the inputs and outputs of the explicit \phs. Thus, under Property~\ref{req:inputs_outputs}, the necessity of \eqref{eq:theo:EF:fullrank} from Proposition~\ref{prop:wellposedness:necc+suff} also accounts for the existence of an explicit \phs.

\section{Main practical result} \label{sec:pracmainresult}
The methods from Sections~\ref{sec:theomainresult} and~\ref{sec:bg2phs} can be summarised in an algorithm which generates an explicit \phs~from a given bond graph. 
This algorithm can be fully automated in a technical computing system and is the main practical result of this paper. 
Algorithm~\ref{algo:bg2phs} presents a program listing which serves as a guide for implementation. 
On the webpage \url{www.irs.kit.edu/2758.php}, we provide an implementation in Wolfram language. 
%
\begin{algorithm}[h!]
	\caption{~} 
	\begin{algorithmic}[1]
		\Statex \textbf{Input:} $\dimBond$-dimensional bond graph
		\ForAll{$i \in \V_\bgI$} 
		\State compute $\F_i \ofx$, $\E_i \ofx$ according to \eqref{eq:patterns:01} or \eqref{eq:patterns:TFGY}
		\State construct $\dirac_i\ofx$ as in \eqref{eq:KDSs}
		\State bring $\dirac_i\ofx$ to the form \eqref{eq:KDS:interiorexterior}
		\State compute $\dirac^\bgIC$ according to \eqref{eq:IC:DS}
		\State compute $\mat{\Gamma}_i^\T \ofx$ according to \eqref{eq:KDS:Mi}
		\EndFor
		\State $\mat{\Gamma}^\T\ofx \leftarrow (\mat{\Gamma}_i^\T\ofx)$ for all $i \in \V_\bgI$
		\State $\mat{\Lambda}^\T\ofx \leftarrow \ker(\mat{\Gamma}^\T\ofx)$
		\State write $\mat{\Lambda}^\T\ofx$ as $(\mat{\Lambda}_i^\T\ofx)$ for all $i \in \V_\bgI$
		\State compute $\dirac \ofx$ according to \eqref{eq:singleKDS}
		\State bring $\dirac \ofx$ to the form \eqref{eq:ds:kernel:blocks}
		\If{\eqref{eq:theo:EF:fullrank} is violated}
		\Print "$\BG$ contains dependent sources"
		\Terminate
		\EndIf
		\If{\eqref{eq:FEF:fullrank} is violated}
		\Print "$\BG$ contains dependent storages or 
		\markcomment{3}{storages determined by sources"}
		\Terminate
		\EndIf
		\State split $\F_{\bgR} \ofx$ such that \eqref{eq:kds2iods:splitting} is fulfilled
		\State split $\E_{\bgR} \ofx$, $\f_{\bgR}$, $\e_{\bgR}$ in same parts as $\F_{\bgR} \ofx$
		\State compute $\mat{Z}\ofx$ according to \eqref{eq:iods:Zuy}
		\State compute $\dirac\ofx$ as in \eqref{eq:dirac:io:specialform}
		\If{\eqref{eq:ass:R-port:theo} does not exist}
		\Print "no suitable input-output splitting of
		\markcomment{3}{\bgR-type elements exists"}
		\Terminate
		\EndIf
		\State bring resistive relation from \eqref{eq:resistivePort} to \eqref{eq:ass:R-port:theo}
		\State compute \phs~matrices with \eqref{eq:real:phs_matrices_from_HIO_DS}
		\State $\vecu \leftarrow \vecu_\bgP$, $\vecy \leftarrow \vecy_\bgP$
		\State \Return explicit \phs~\eqref{eq:phs}
	\end{algorithmic}
	\label{algo:bg2phs}
\end{algorithm}



\section{Academic example} \label{sec:example}
\begin{figure}[htb]
	\begin{center}
		\includegraphics[width=0.45\textwidth]{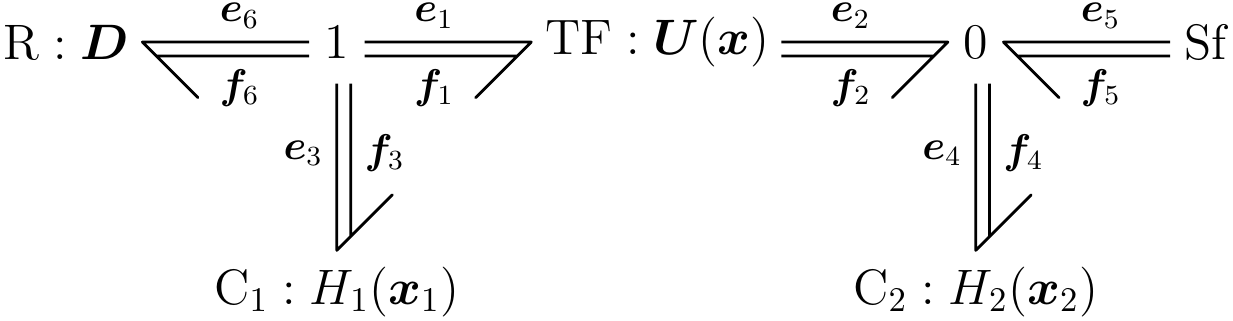}
		\caption[Exemplary bond graph]{Exemplary $\dimBond$-dimensional bond graph.}
		\label{fig:example}
	\end{center}
\end{figure}

In this section, we illustrate the main theoretical and practical results of this paper through an academic example. 
Consider the $\dimBond$-dimensional bond graph in Figure~\ref{fig:example}. 
The elements and bonds are summarised in the sets $\V=\V_\bgI \cup\V_\bgE$, $\B=\B_\bgI \cup\B_\bgE$, respectively, with $\V_\bgI = \{\bgZero,\bgOne,\bgTF\}$, $\V_\bgE =\{\bgC_1,\bgC_2,\bgR,\bgSf\}$, $\B_\bgI=\{1,2\}$, and $\B_\bgE=\{3,4,5,6\}$. 
The system state vector is $\x=\left(\x_1^\T\,\x_2^\T\right)^\T\in\R^{2\dimBond}$. We suppose an arbitrary, differentiable, non-negative storage function $H\ofx = H_1(\x_1) + H_2(\x_2)$. 
The \bgR-type element is specified by a matrix $\mat{D}\in\R^{\dimBond \times \dimBond}$ with $\mat{D}=\mat{D}^\T\posdef$. 
The transformer $\bgTF$ is state-modulated with full rank matrix $\mat{U}\ofx=\mat{U}^\T\ofx\in\R^{\dimBond \times \dimBond}$. 
\\
First, for each element $i\in\V_\bgI$ we formulate a Dirac structure $\dirac_i$ of the form~\eqref{eq:KDSs}. With the matrices from \eqref{eq:FE:0}, \eqref{eq:FE:1}, and \eqref{eq:FE:TF}, the equation systems of the Dirac structures are:
\begin{subequations} \label{eq:ex:ds:kernel}
\begin{alignat}{2}
&\dirac_\bgZero:  
&&\begin{pmatrix} 
\blockI&\blockI&\blockI \\  
\blockZero&\blockZero&\blockZero\\ 
\blockZero&\blockZero&\blockZero
\end{pmatrix} 
\begin{pmatrix}
\f_{5}\\
\f_{2}\\
-\f_{4}
\end{pmatrix} +
\begin{pmatrix}
\blockZero&\blockZero&\blockZero\\
\blockI&-\blockI&\blockZero\\
\blockI&\blockZero&-\blockI
\end{pmatrix}
\begin{pmatrix}
\e_{5}\\
\e_{2}\\
\e_{4}
\end{pmatrix}
=\0, \label{eq:ex:ds:kernel:0}\\
&\dirac_\bgOne: 
&&\begin{pmatrix} 
\blockZero&\blockZero&\blockZero\\ 
-\blockI&\blockI&\blockZero\\  
-\blockI&\blockZero&\blockI
\end{pmatrix} 
\begin{pmatrix}
-\f_{3}\\
-\f_{6}\\
-\f_{1}
\end{pmatrix} +
\begin{pmatrix}
-\blockI&-\blockI&-\blockI \\  
\blockZero&\blockZero&\blockZero\\
\blockZero&\blockZero&\blockZero
\end{pmatrix}
\begin{pmatrix}
\e_{3}\\
\e_{6}\\
\e_{1}
\end{pmatrix}
=\0, \label{eq:ex:ds:kernel:1}\\
&\dirac_\bgTF: 
&&\begin{pmatrix} 
\blockI&\mat{U}\ofx\\  
\blockZero&\blockZero
\end{pmatrix} 
\begin{pmatrix}
\f_{1}\\
-\f_{2}
\end{pmatrix}+
\begin{pmatrix}
\blockZero&\blockZero\\
-\mat{U}\ofx&\blockI
\end{pmatrix}
\begin{pmatrix}
\e_{1}\\
\e_{2}
\end{pmatrix}
=\0. \label{eq:ex:ds:kernel:TF}
\end{alignat}
\end{subequations}
Throughout this example, square zero matrices and identity matrices are of dimension $\dimBond$. 
Rewrite \eqref{eq:ex:ds:kernel:0} as 
\begin{equation}
\begin{pmatrix} 
\blockI&\blockI&\blockI \\  
\blockZero&\blockZero&\blockZero\\ 
\blockZero&\blockZero&\blockZero
\end{pmatrix} 
\begin{pmatrix}
-\f_{4}\\
\f_{5}\\
\f_{2}
\end{pmatrix} +
\begin{pmatrix}
\blockZero&\blockZero&\blockZero\\
\blockZero&\blockI&-\blockI\\
-\blockI&\blockI&\blockZero
\end{pmatrix}
\begin{pmatrix}
\e_{4}\\
\e_{5}\\
\e_{2}
\end{pmatrix}
=\0. \label{eq:ex:ds:kernel:0:rewritten}
\end{equation}
Equations~\eqref{eq:ex:ds:kernel:1}, \eqref{eq:ex:ds:kernel:TF}, \eqref{eq:ex:ds:kernel:0:rewritten} are then of the form~\eqref{eq:KDS:interiorexterior}. The equation system of the interconnection Dirac structure $\dirac_\bgIC$ from \eqref{eq:IC:DS} is given by: 
\begin{equation}\label{eq:ex:dsIC}
	\underbrace{\left(
		\begin{array}{c}
		\blockZero\\
		\blockI\\
		\blockZero\\
		\blockZero
		\end{array}\right.}_{\F^\bgIC_\bgZero}
	\underbrace{\left.
		\begin{array}{c}
		\blockI\\
		\blockZero\\
		\blockZero\\
		\blockZero
		\end{array}\right.}_{\F^\bgIC_\bgOne}
	\underbrace{\left.
		\begin{array}{cc}
		\blockI&\blockZero\\
		\blockZero&\blockI\\
		\blockZero&\blockZero\\
		\blockZero&\blockZero
		\end{array}\right)}_{\F^\bgIC_\bgTF} \!\!
	\begin{pmatrix} 
	\f_{2}\\
	-\f_{1}\\
	\f_{1}\\
	-\f_{2}
	\end{pmatrix}
	+ 
	\underbrace{\left(
		\begin{array}{c}
		\blockZero\\
		\blockZero\\
		\blockZero\\
		\blockI
		\end{array}\right.}_{\E^\bgIC_\bgZero}
	\underbrace{\left.
		\begin{array}{c}
		\blockZero\\
		\blockZero\\
		-\blockI\\
		\blockZero
		\end{array}\right.}_{\E^\bgIC_\bgOne}
	\underbrace{\left.
		\begin{array}{cc}
		\blockZero&\blockZero\\
		\blockZero&\blockZero\\
		\blockI&\blockZero\\
		\blockZero&-\blockI
		\end{array}\right)}_{\E^\bgIC_\bgTF} \!\!
	\begin{pmatrix} 
	\e_{2}\\
	\e_{1}\\
	\e_{1}\\
	\e_{2}
	\end{pmatrix} 
	  =  \0.
\end{equation} 
With \eqref{eq:ex:dsIC}, we obtain 
\begin{equation}
\mat{\Gamma}^\T \ofx =\underbrace{\left(
	\begin{array}{ccc}
	\blockZero&\blockZero&\blockZero\\
	\blockZero&-\blockI&\blockZero\\
	\blockZero&\blockZero&\blockZero\\
	\blockI&\blockZero&\blockZero
	\end{array}\right.}_{\mat{\Gamma}_{\bgZero}}
\underbrace{\left.
	\begin{array}{cccc}
	-\blockI&\blockZero&\blockZero\\
	\blockZero&\blockZero&\blockZero\\
	\blockZero&\blockZero&-\blockI\\
	\blockZero&\blockZero&\blockZero
	\end{array}\right.}_{\mat{\Gamma}_{\bgOne}}
\underbrace{\left.
	\begin{array}{cc}
	\blockZero&-\mat{U}\ofx\\
	\blockZero&\blockI\\
	\blockI&\blockZero\\
	-\mat{U}\ofx&\blockZero
	\end{array}\right)}_{\mat{\Gamma}_{\bgTF}\ofx}.
\end{equation}
A matrix $\mat{\Lambda}\ofx $ with $\img(\mat{\Lambda}^\T\ofx )=\ker(\mat{\Gamma}^\T\ofx )$ for all $\x\in\X$ is given by
\begin{equation} \label{eq:ex:Lambda}
\mat{\Lambda}\ofx =\underbrace{\left(
	\begin{array}{ccc}
	\blockZero&\blockI&\blockZero\\
	\mat{U}\ofx&\blockZero&\blockZero\\
	\blockZero&\blockZero&\blockZero\\
	\blockZero&\blockZero&\blockI
	\end{array}\right.}_{\mat{\Lambda}_{\bgZero}\ofx}
\underbrace{\left.
	\begin{array}{cccc}
	-\mat{U}\ofx&\blockZero&\blockZero\\
	\blockZero&\blockZero&\blockI\\
	\blockZero&\blockI&\blockZero\\
	\blockZero&\blockZero&\blockZero
	\end{array}\right.}_{\mat{\Lambda}_{\bgOne}\ofx}
\underbrace{\left.
	\begin{array}{cc}
	\blockZero&\blockI\\
	\blockI&\blockZero\\
	\blockZero&\blockZero\\
	\blockZero&\blockZero\\
	\end{array}\right)}_{\mat{\Lambda}_{\bgTF}}.
\end{equation}
With \eqref{eq:ex:Lambda}, we can compute a single Dirac structure describing the equations of the junction structure. The equation system of the composed Dirac structure is
\begin{multline} \label{eq:ex:ds:implicit}
	\underbrace{\left(
	\begin{array}{cc}
	\blockZero&\blockZero\\
	-\blockI& \mat{U}\ofx\\
	-\blockI&\blockZero\\
	\blockZero&\blockZero
	\end{array}\right.}_{\F_\bgC(\x)}
	\underbrace{\left.
	\begin{array}{c}
	\blockZero\\
	\blockZero\\
	\blockI\\
	\blockZero
	\end{array}\right.}_{\F_\bgR}
	\underbrace{\left.
	\begin{array}{c}
	\blockZero\\
	 \mat{U}\ofx\\
	\blockZero\\
	\blockZero
	\end{array}\right)}_{\F_\bgSf(\x)}  
	\begin{pmatrix}
	-\f_{3}\\
	-\f_{4}\\
	-\f_{6}\\
	\f_{5}
	\end{pmatrix}
	+ \\
	\underbrace{\left(
	\begin{array}{cc}
	\mat{U}\ofx&\blockZero\\
	\blockZero&\blockZero\\
	\blockZero&\blockZero\\
	\blockZero&-\blockI
	\end{array}\right.}_{\E_\bgC(\x)}
	\underbrace{\left.
	\begin{array}{cc}
	 \mat{U}\ofx\\
	\blockZero\\
	\blockZero\\
	\blockZero
	\end{array}\right.}_{\E_\bgR(\x)}
	\underbrace{\left.
	\begin{array}{c}
	\blockI\\
	\blockZero\\
	\blockZero\\
	\blockI
	\end{array}\right)}_{\E_\bgSf}
\begin{pmatrix}
	\e_{3}\\
	\e_{4}\\
	\e_{6}\\
	\e_{5}
	\end{pmatrix} 
	  =  \0.
\end{multline} 
Since $\rank\left(\F_\bgC\ofx~\E_\bgSf\right)=3\dimBond$ for all $\x\in\R^{2\dimBond}$, Assumptions~\ref{ass:independentsources} and \ref{ass:storagesandsources:independent} are fulfilled. Using Corollary~\ref{cor:kds2iods:specialform} with $\F_{\bgR,2} = \F_{\bgR}$ we obtain an explicit representation of \eqref{eq:ex:ds:implicit}
\begin{equation} \label{eq:ex:ds:explicit}	
	\begin{pmatrix}
		-\f_{3}\\
		-\f_{4}\\
		\e_{6}\\		
		\e_{5}
	\end{pmatrix}
	=
	\left(
	\begin{array}{cc;{2pt/2pt}c;{2pt/2pt}c}
	\blockZero&\blockZero&\blockI&\blockZero\\
	\blockZero&\blockZero&\mat{V}\ofx&-\blockI\\ \hdashline[2pt/2pt]
	-\blockI&-\mat{V}\ofx&\blockZero&\blockZero\\ \hdashline[2pt/2pt]
	\blockZero&\blockI&\blockZero&\blockZero
	\end{array}
	\right)
	\begin{pmatrix}
		\e_{3}\\
		\e_{4}\\
		-\f_{6}\\		
		\f_{5}
	\end{pmatrix}
\end{equation}
where $\mat{V}\ofx=\mat{U}^{-1}\ofx$. The dashed lines indicate the matrix blocks according to \eqref{eq:dirac:io:specialform}. 
For an input-output splitting of the \bgR-type element as in \eqref{eq:ex:ds:explicit}, Assumption~\ref{ass:resistiveRelation} is satisfied. 
With \eqref{eq:real:phs_matrices_from_HIO_DS} we then obtain the following explicit \phs 
\begin{subequations} \label{eq:ex:phs}
\begin{alignat}{2}
	\dot{\x} &=
	- &&\underbrace{	\begin{pmatrix}
		\mat{D}&\mat{D}\mat{V}\ofx\\
		\mat{V}\ofx \!\mat{D}&~\mat{V}\ofx \! \mat{D}\mat{V}\ofx
		\end{pmatrix}}_{=\matR\,\ofx} 
		\frac{\partial H}{\partial \x} \ofx
		+
		\underbrace{
		\begin{pmatrix}
		\blockZero \\ \blockI
		\end{pmatrix}}_{=\matG}
		\f_{5}, 
		\\
		\e_{5} & = 
		&&\underbrace{\begin{pmatrix}
			\blockZero&&\blockI
		\end{pmatrix}}_{=\matG^\T} \frac{\partial H}{\partial \x} \ofx,
\end{alignat}
\end{subequations}
with $\matJ\ofx$, $\matP\ofx$, $\matM\ofx$, $\matS\ofx$ being zero. By the properties of $\mat{D}$ and $\mat{U}\ofx$ we indeed have $\matR\ofx=\matR^\T\ofx$ for all $x\in\R^{2\dimBond}$. Moreover, \eqref{eq:ex:phs} has Property~\ref{req:inputs_outputs}. 	

\section{Conclusion} \label{sec:conclusion}
In this paper, we present a method for an explicit port-Hamiltonian formulation of multi-bond graphs (Theorem~\ref{theorem:main_result}). 
Furthermore, we provide two conditions for the existence of such an explicit formulation -- one necessary and one sufficient.  
The method can be fully automated (Algorithm~\ref{algo:bg2phs}); along with this publication, we provide an implementation in Wolfram language. 
Future work will address the generalisation of our modelling method to multi-bond graphs containing dependent storages and/or storages determined by sources, i.e. the case where \eqref{eq:FEF:fullrank} is violated. 


\bibliographystyle{elsart-num}       
\bibliography{./include/bibMartin,./include/bibPHS,./include/bibMotivation}           

\begin{thebibliography}{10}
\expandafter\ifx\csname url\endcsname\relax
  \def\url#1{\texttt{#1}}\fi
\expandafter\ifx\csname urlprefix\endcsname\relax\def\urlprefix{URL }\fi

\bibitem{maschke1992a}
B.~Maschke, A.~{van der Schaft}, Port-controlled {H}amiltonian systems:
  modelling origins and systemtheoretic properties, IFAC Proceedings Volumes
  25~(13) (1992) 359--365.

\bibitem{maschke1992b}
B.~Maschke, A.~{van der Schaft}, P.~C. Breedveld, An intrinsic {H}amiltonian
  formulation of network dynamics: Non-standard poisson structures and
  gyrators, Journal of the Franklin Institute 329~(5) (1992) 923--966.

\bibitem{vanderSchaft1995}
A.~{van der Schaft}, B.~Maschke, The {Hamiltonian} formulation of energy
  conserving physical systems with external ports, AEU - Archiv f{\"u}r
  Elektronik und {\"U}bertragungstechnik 49~(5-6) (1995) 362--371.

\bibitem{mehl2016}
C.~Mehl, V.~Mehrmann, P.~Sharma, Stability radii for linear {Hamiltonian}
  systems with dissipation under structure-preserving perturbations, SIAM
  Journal on Matrix Analysis and Applications 37~(4) (2016) 1625--1654.

\bibitem{kotyczka2018}
P.~Kotyczka, L.~Lef{\`e}vre, Discrete-time port-{Hamiltonian} systems based on
  {G}auss-{L}egendre collocation, IFAC-PapersOnLine 51~(3) (2018) 125--130.

\bibitem{vanderSchaft2002}
A.~{van der Schaft}, B.~Maschke, {Hamiltonian} formulation of
  distributed-parameter systems with boundary energy flow, Journal of Geometry
  and Physics 42~(1-2) (2002) 166--194.

\bibitem{legorrec2005}
Y.~Le~Gorrec, H.~Zwart, B.~Maschke, Dirac structures and boundary control
  systems associated with skew-symmetric differential operators, SIAM Journal
  on Control and Optimization 44~(5) (2005) 1864--1892.

\bibitem{jacob2012}
B.~Jacob, H.~Zwart, Linear port-{Hamiltonian} systems on infinite-dimensional
  spaces, Vol. 223, Springer Science \& Business Media, 2012.

\bibitem{ramirez2017}
H.~Ramirez, H.~Zwart, Y.~Le~Gorrec, Stabilization of infinite dimensional
  port-{Hamiltonian} systems by nonlinear dynamic boundary control, Automatica
  85 (2017) 61--69.

\bibitem{kugi2001}
A.~Kugi, Non-linear control based on physical models: electrical, mechanical
  and hydraulic systems, Lecture {Notes} in {Control} and {Information}
  {Sciences}, Springer, London, 2001.

\bibitem{duindam2009}
V.~Duindam, A.~Macchelli, S.~Stramigioli, H.~Bruyninckx (Eds.), Modeling and
  control of complex physical systems: the port-{Hamiltonian} approach,
  Springer, Berlin Heidelberg, 2009.

\bibitem{kotyczka2013}
P.~Kotyczka, Local linear dynamics assignment in {IDA}-{PBC}, Automatica 49~(4)
  (2013) 1037--1044.

\bibitem{ortega2008}
R.~Ortega, A.~{van der Schaft}, F.~Castanos, A.~Astolfi, Control by
  interconnection and standard passivity-based control of port-{Hamiltonian}
  systems, IEEE Transactions on Automatic control 53~(11) (2008) 2527--2542.

\bibitem{doerfler2009}
F.~D{\"o}rfler, J.~K. Johnsen, F.~Allg{\"o}wer, An introduction to
  interconnection and damping assignment passivity-based control in process
  engineering, Journal of Process Control 19~(9) (2009) 1413--1426.

\bibitem{venkatraman2010}
A.~Venkatraman, A.~{van der Schaft}, Full-order observer design for a class of
  port-{Hamiltonian} systems, Automatica 46~(3) (2010) 555--561.

\bibitem{vincent2016}
B.~Vincent, N.~Hudon, L.~Lef\`{e}vre, D.~Dochain, Port-{Hamiltonian} observer
  design for plasma profile estimation in tokamaks, IFAC-PapersOnLine 49~(24)
  (2016) 93--98.

\bibitem{van_der_schaft2016}
A.~van~der Schaft, L2-Gain and Passivity Techniques in Nonlinear Control, 3rd
  Edition, Communications and {Control} {Engineering}, Springer, Cham,
  Switzerland, 2016.

\bibitem{vanderSchaftMaschke2013}
A.~van~der Schaft, B.~Maschke, Port-{Hamiltonian} systems on graphs, SIAM
  Journal on Control and Optimization 51~(2) (2013) 906--937.

\bibitem{falaize2016}
A.~Falaize, T.~H{\'{e}}lie, Passive guaranteed simulation of analog audio
  circuits: a port-{Hamiltonian} approach, Applied Sciences 6~(10).

\bibitem{pyphs2019}
A.~Falaize, T.~H\'{e}lie, {PyPHS}, \url{https://github.com/pyphs/pyphs},
  accessed: 2019-07-03.

\bibitem{rosenberg1971}
R.~C. Rosenberg, State-space formulation for bond graph models of multiport
  systems, Journal of Dynamic Systems, Measurement, and Control 93~(1) (1971)
  35--40.

\bibitem{wellstead1979}
P.~E. Wellstead, Introduction to physical system modelling, Acadamic Press,
  London, 1979.

\bibitem{golo2003}
G.~Golo, A.~van~der Schaft, P.~Breedveld, B.~Maschke, {Hamiltonian} formulation
  of bond graphs, in: Nonlinear and hybrid systems in automotive control,
  Springer, London, 2003, pp. 351--372.

\bibitem{lopes2016}
N.~Lopes, Approche passive pour la mod\'{e}lisation, la simulation et
  l'\'{e}tude d'un banc de test robotis\'{e} pour les instruments de type
  cuivre, Ph.D. thesis, Universit\'{e} Pierre et Marie Curie, Paris (2016).

\bibitem{donaire2009}
A.~Donaire, S.~Junco, Derivation of input-state-output port-{Hamiltonian}
  systems from bond graphs, Simulation Modelling Practice and Theory 17~(1)
  (2009) 137--151.

\bibitem{dai2016}
S.~Dai, Compositional modeling and design of cyber-physical systems using
  port-{Hamiltonian} systems, Ph.D. thesis, Vanderbilt University, Nashville,
  Tennessee (2016).

\bibitem{borutzky2010}
W.~Borutzky, Bond Graph Methodology: Development and Analysis of
  Multidisciplinary Dynamic System Models, Springer, London, 2010.

\bibitem{van_der_schaft2014}
A.~{van der Schaft}, D.~Jeltsema, Port-{Hamiltonian} systems theory: An
  introductory overview, Foundations and Trends\textregistered~in Systems and
  Control 1~(2-3) (2014) 173--378.

\bibitem{karnopp2012}
D.~C. Karnopp, D.~L. Margolis, R.~C. Rosenberg, System Dynamics: Modeling,
  Simulation, and Control of Mechatronic Systems, 5th Edition, John Wiley \&
  Sons, Incorporated, Hoboken, USA, 2012.

\bibitem{batlle2011}
C.~Batlle, I.~Massana, E.~Sim\'{o}, Representation of a general composition of
  {Dirac} structures, in: 2011 50th {IEEE} {Conference} on {Decision} and
  {Control} and {European} {Control} {Conference}, 2011, pp. 5199--5204.

\bibitem{bloch1999}
A.~Bloch, P.~E. Crouch, Representations of {Dirac} structures on vector spaces
  and nonlinear {L}-{C} circuits, in: Differential {Geometry} and {Control},
  Vol.~64 of Symposia in {Pure} {Mathematics}, American Mathematical Society,
  Providence, Rhode Island, 1999, pp. 103--117.

\bibitem{golo2000}
G.~Golo, P.~Breedveld, B.~Maschke, A.~van~der Schaft, Geometric formulation of
  generalized bond graph models - part i: Generalized junction structures,
  Technical {Report}, University of Twente, Faculty of Mathematical Sciences,
  Enschede (2000).

\end{thebibliography}



\begin{appendix}
\section{Dirac structures} \label{sec:preliminaries}
In Appendix~\ref{sec:preliminaries} we recapitulate some representations of Dirac structures. For a detailed introduction, we refer the reader to \cite{van_der_schaft2014}.  \\
Given an abstract finite-dimensional vector space $\vecspace{F}$ and its dual vector space $\vecspace{E}\define\vecspace{F}^*$. The spaces $\vecspace{F}$ and $\vecspace{E}$ are referred to as \emph{space of flows} and \emph{space of efforts}, respectively. We denote $\f \in \vecspace{F}$ as \emph{flow vectors} and $\e \in \vecspace{E}$ as \emph{effort vectors}. 
\begin{defn}[\cite{van_der_schaft2014}]\label{def:DS}
A subspace $\dirac \subset \vecspace{F} \times \vecspace{E}$ is a \emph{constant Dirac structure} if
\begin{subequations}
\label{eq:dirac:def}
		\begin{alignat}{2}
		&(i)	\quad	&& \langle \e \mid \f \rangle = 0, \quad\forall \left(\f,\e \right) \in \dirac, \label{eq:dirac:def:i}\\
		&(ii)	\quad	&& \dim \dirac = \dim \vecspace{F} \label{eq:dirac:def:ii},
		\end{alignat}
\end{subequations}
where $\langle \e \mid \f \rangle = \e(\f)$ denotes the dual pairing.
\end{defn}
\begin{rem}
Throughout this paper we have $\vecspace{F}=\R^n$. As $\vecspace{E} = \left(\R^n \right)^* \isomorph \R^n$, we identify $\vecspace{E}$ with $\R^n$.
\end{rem}
%
%
\begin{defn}\label{def:DS:modulated}
A \emph{modulated Dirac structure} is a family of constant Dirac structures  $\dirac \ofx \subset \R^n \times \R^n$ indexed over $\x \in \vecspace{X}$.
\end{defn}
%
%
\begin{defn}\label{def:ds:kernel}
A \emph{kernel representation} of a modulated Dirac structure $\dirac \ofx \subset \R^n \times \R^n$ with $\x \in \vecspace{X}$ is 
\begin{align} \label{eq:ds:kernel}
	\dirac\ofx \tighteq\lbrace
	(\f,\e) \in \R^n \! \times \! \R^n \mid \F\ofx \! \f+\E\ofx \! \e = \vec{0}\rbrace,
\end{align}
%
%
where the matrices $\F\ofx$ and $\E\ofx$ satisfy
\begin{subequations}
\label{eq:dirac:kernel:cond}
		\begin{alignat}{2}\label{eq:diracKern}
		&(i)\quad &&\E\ofx\F\Tofx+\F\ofx\E\Tofx=\0,\\
		&(ii)\quad&&\rank (\F\ofx~\E\ofx)=n  \label{eq:diracRank}
		\end{alignat}
\end{subequations}
for all $\x \in \vecspace{X}$.

\begin{rem} \label{rem:uniqueness:FE}
The matrices \(\F\ofx\) and \(\E\ofx\) are not uniquely determined by the kernel representation. For example both matrices can be multiplied from the left by an arbitrary invertible matrix \(\mat{T}\ofx\) without changing \(\dirac\).
\end{rem}
%
%
\end{defn}
\begin{defn} \label{def:DS:io}
Let $\dirac \ofx \subset \R^{n} \times \R^{n}$ with $\x \in \vecspace{X}$ be a modulated Dirac structure and $\left(\f,\e\right)\in\dirac\ofx$. Possibly after permutations split $\f$ into $\left(\f_1^\T ~ \f_2^\T \right)^\T$. Correspondingly, split $\e$ into $\left(\e_1^\T ~ \e_2^\T \right)^\T$. An \emph{input-output representation} of $\dirac\ofx$ is
\begin{align} \label{eq:ds:io}
	\dirac\ofx =\lbrace
\left( \f,\e \right)
	\in \R^n \times \R^n \mid 
	\vecy
	= \matioD \ofx 
		\vecu
	\rbrace.
\end{align} 
where $\vecu = \left(\e_1^\T ~ \f_2^\T \right)^\T$ and $\vecy = \left(\f_1^\T ~ \e_2^\T \right)^\T$ are referred to as \emph{input vector} and \emph{output vector}, respectively. The matrix $\matioD\ofx$ satisfies $\matioD\ofx = -\matioD \ofx ^\T$ for all $\x \in \vecspace{X}$. 
%
\end{defn}
\begin{rem} \label{rem:DS:imex}
Due to the structure of the equation systems, we denote \eqref{eq:ds:kernel} and \eqref{eq:ds:io} as \emph{implicit} and \emph{explicit} representations, respectively. 
\end{rem}

\end{appendix}

\listoffixmes
\end{document}